\documentclass[a4paper,10pt]{article}
\usepackage[utf8]{inputenc}
\usepackage{amsmath}
\allowdisplaybreaks[0]
\usepackage[ngerman,british]{babel}
\usepackage{hyperref}
\usepackage[left=30mm, right=30mm, top=25mm, bottom=25mm,bindingoffset=5mm, twoside]{geometry}
\usepackage{graphicx}
\usepackage{relsize}
\usepackage{cite}
\usepackage{slashed}
\usepackage[textwidth=1.5cm]{todonotes}

\newcommand{\Delrho}[2][{}]{\Delta\rho_\text{#1}^\text{(#2)}}
\newcommand{\delrho}[2][{}]{\delta\rho_\text{#1}^\text{(#2)}}
\newcommand{\Delr}[2]{\Delta r^\text{#1}_\text{#2}}
\newcommand{\Dalpha}{\Delta\alpha}
\newcommand{\MhO}{m_{h^0}}
\newcommand{\MHp}{m_{H^\pm}}
\newcommand{\MHH}{m_{H^0}}
\newcommand{\MAO}{m_{A^0}}
\newcommand{\dMVsq}[2]{\delta^{(#1)}\!M_{#2}^2}
\newcommand{\dMWsq}[1][1]{\delta^{(#1)}\!M_W^2}
\newcommand{\dMZsq}[1][1]{\delta^{(#1)}\!M_Z^2}

\newcommand{\musq}{\mu_2^2}
\newcommand{\dSWsq}[1][1]{\delta^{(#1)} s_W^2}
\newcommand{\dCWsq}[1][1]{\delta^{(#1)} c_W^2}

\newcommand{\dZe}[1]{\delta^\text{(#1)}\! Z_e}

\newcommand{\Self}[3][\text{}]{\Sigma^{#2}_{#1}\left(#3\right)}

\newcommand{\PiAA}[2][\text{}]{\Pi^{\gamma}_{#1}\left(#2\right)}

\newcommand{\SB}{s_\beta}
\newcommand{\CB}{c_\beta}
\newcommand{\TB}{t_\beta}

\newcommand{\lfive}{\lambda_5}

\renewcommand{\Re}{\operatorname{Re}}

\newcommand{\order}[1]{\mathcal{O}\left(#1\right)}
\newcommand{\Rot}[1]{\mathbf{R}\left(#1\right)}

\newcommand{\LIHDM}{\Lambda_{345}}

\newcommand{\DelK}[2][{}]{\Delta\kappa_\text{#1}^\text{(#2)}}
\newcommand{\FV}[1]{F_V^{#1}\left(M_Z^2\right)}
\newcommand{\FA}[1]{F_A^{#1}\left(M_Z^2\right)}
\newcommand{\DelrhoG}[2][{}]{\Delta\overline{\rho}_\text{#1}^\text{(#2)}}

\newcommand{\mwsm}{M_{W,\text{SM}}}
\newcommand{\mmwsm}{M_{W,\text{SM}}^2}
\newcommand{\slsm}{s^2_{l,\text{SM}}}

\begin{document}
\thispagestyle{empty}

\hfill  
\begin{large} 
MPP-2022-66
\end{large}

\vspace*{2.5cm}

\vspace{0.5cm}

\begin{center}

\begin{Large}
\textbf{Two-loop improved predictions 
for $\mathbf{M_W}$ and $\mathbf{sin^2\theta_{eff}}$ \\[0.2cm]
in Two-Higgs-Doublet Models} 

\end{Large}

\vspace*{1.5cm}

\begin{large}

{\sc Stephan Hessenberger}\footnote{email: s.hessenberger@gmx.de}  
and {\sc Wolfgang Hollik}\footnote{email: hollik@mpp.mpg.de}  

\vspace*{0.7cm}
   {\it Max Planck Institut f\"ur Physik \\[0.1cm] 
         (Werner-Heisenberg-Institut) \\[0.1cm] 
         F\"ohringer Ring 6,  80805 M\"unchen, Germany }  

\end{large}

\end{center}

\vspace*{1.5cm}

\begin{large}

\begin{abstract}
We present the currently most precise predictions for the 
$W$-boson mass 
and the leptonic effective mixing angle 
in the aligned Two-Higgs-Doublet Model.
The evaluation includes the full one-loop result, 
all known higher-order corrections of the 
Standard Model, and the non-standard two-loop contributions
that increase with mass splittings between charged and neutral
THDM Higgs bosons.
They depend on $\tan\beta$ and the soft $Z_2$-symmetry breaking
parameter $m_{12}^2$ of the scalar potential, 
in addition to the non-standard boson masses. Whenever the
one-loop corrections become large, the two-loop contributions
yield substantial modifications of the predictions, which is of
particular importance for the $W$ mass  where large mass shifts 
are required to reach the recently published final result
of the CDF collaboration.
Numerical results are shown for  the dependence on the various
non-standard parameters and in comparison with experimental data.  
\end{abstract}

\end{large}


\clearpage

\section{Introduction}

Precise measurements of electroweak observables at lepton and hadron colliders
provide important tests of the electroweak Standard Model (SM) as well as of possible physics
 beyond the Standard Model (BSM).
Model parameters entering the theoretical predictions via loop corrections can be constrained by  
comparison with the measured data provided the theoretical precision can compete with
the experimental accuracy. 
Prominent cases of precision observables, measured with highest accuracy and
very sensitive to virtual heavy particles, are the effective leptonic mixing angle
in terms of $\sin^2\theta_{\rm eff}$, and the correlation of the 
gauge boson masses $M_W,M_Z$ via the Fermi constant $G_F$.
The latter allows a precise prediction of $M_W$ from $M_Z$, the electromagnetic 
fine-structure constant and the Fermi constant, in combination with additional parameters 
in the higher-order contributions.
With the discovery of a SM-like Higgs boson by the LHC experiments
ATLAS~\cite{Aad:2012tfa} and CMS~\cite{Chatrchyan:2012ufa} 
all the required input parameters of the SM are determined, and together with the adequate
calculation of the loop corrections, acccurate predictions for the electroweak observables 
have become available.

Measurements of $M_W$ at LEP~\cite{Schael:2013ita}, Tevatron~\cite{Aaltonen:2013iut}, 
and the LHC~\cite{ATLAS:2017rzl}
resulted in the current world average~\cite{ParticleDataGroup:2020ssz}
\begin{equation}
 M_{W,\text{exp}}=80.379\pm 0.012\text{ GeV}.
\label{Eq:Wexp}
\end{equation}
Recently the final result of the $W$ mass measurement by the  
CDF Collaboration~\cite{CDF:2022hxs} has been presented, 
 \begin{equation}
 M_{W,\text{CDF}}=80.4335\pm 0.0094\text{ GeV} 
\label{Eq:WCDF}
\end{equation}
which is in obvious tension with the average~(\ref{Eq:Wexp})
from previous measurements and with the SM prediction.

The effective leptonic mixing angle at the $Z$ resonance has been measured in 
electron--positron reactions $e^+e^- \to f\bar{f}$ at the colliders LEP and SLC  
via the forward--backward and $\tau$-polarization asymmetries
as well as the left--right asymmetry
accessible by longitudinally polarized initial-state electrons that were available at the SLC.
The average is given by~\cite{ALEPH:2005ab}
\begin{equation}
 \sin^2\theta_{\rm eff,exp} =0.23153\pm 0.00016.
 \label{Eq:slep_exp}
\end{equation}
It is worth to note that the individual value measured by the
SLD Collaboration~\cite{SLD:2000leq} via the left--right asymmetry,
\begin{equation}
 \sin^2\theta_{\rm eff,SLD} =0.23097\pm 0.00027 ,
 \label{Eq:slep_SLD}
\end{equation}
is about two standard deviations lower than~(\ref{Eq:slep_exp}) 
where all asymmetries at the $Z$ resonance have been combined.  
It is the most precise measurement of the mixing angle from a single observable.

The SM prediction of $M_W$ is based on the complete one-loop~\cite{Sirlin:1980,Marciano:1980pb} and two-loop contributions\cite{Djouadi:1987,Djouadi:1988,Consoli:1989fg,Kniehl:1990,Halzen:1991,Kniehl:1992,Djouadi:1993ss,Freitas:2000gg,Freitas:2002ja,Awramik:2003ee,Awramik:2002wn,Onishchenko:2002ve,Awramik:2002vu,Degrassi:2014sxa} and has been further improved by the leading higher-order corrections up to four 
loops~\cite{Avdeev:1994db,Chetyrkin:1995ix,Chetyrkin:1995js,Chetyrkin:1996cf,Faisst:2003px,vanderBij:2000cg,Schroder:2005db,Chetyrkin:2006bj,Boughezal:2006xk}.
The same level of accuracy has been achieved for the effective leptonic mixing 
angle~\cite{Gambino:1994,Degrassi:1996ps,Djouadi:1987,Djouadi:1988,Kniehl:1990,Halzen:1991,Kniehl:1992,Djouadi:1993ss,Awramik:2004ge,Hollik:2005va,Hollik:2005ns,Hollik:2006ma,Awramik:2006ar,Awramik:2006uz}.

For testing extensions of the SM the same level of accuracy is desirable. 
For example, in the minimal supersymmetric version of the Standard Model, 
the MSSM,  the one-loop  results \cite{Barbieri:1983wy,Grifols:1983gu,Lim:1983re,Eliasson:1984yu,Hioki:1985wz,Grifols:1984xs,Barbieri:1989dc,Drees:1990dx,Drees:1991zk,Chankowski:1993eu,Chankowski:1992er,Garcia:1993sb,Dabelstein:1995ui,Pierce:1996zz,Heinemeyer:2006px} 
were improved by including the leading two-loop corrections via calculations of the contributions
$\Delta\rho$ to the $\rho$-parameter of $\order{\alpha\alpha_s}$
\cite{Djouadi:1996pa,Djouadi:1998sq}
and of $\order{\alpha_t^2}$, $\order{\alpha_t\alpha_b}$, $\order{\alpha_b^2}$ 
\cite{Heinemeyer:2002jq,Haestier:2005ja}. 
For a detailed analysis of precision observables in the MSSM see for example \cite{Heinemeyer:2004gx,Heinemeyer:2006px,Heinemeyer:2007bw,Heinemeyer:2013dia,Stal:2015zca}.

The mere extension of the SM by a second Higgs doublet, the Two-Higgs-Doublet Model (THDM),  
can provide a new type of large contributions to $\Delta\rho$ arising fom the Higgs sector, 
in contrast to the SM where the Higgs sector respects the custodial symmetry and hence 
deviations of $\rho$ from unity can only originate from other sources, like gauge and 
Yukawa couplings. The THDM Higgs sector accommodates three neutral and a pair of 
charged Higgs particles, with masses as independent free parameters; in general,
custodial symmetry is broken and thus $\rho$~different from unity can occur for mass 
splittings between different isospin states of the scalar sector.
The non-standard one-loop corrections to precision observables
are very sensitive to mass differences between charged and neutral Higgs bosons 
\cite{Bertolini:1985ia,Hollik:1986gg,Hollik:1987fg,Denner:1991ie,Froggatt:1991qw,Chankowski:1999ta,Grimus:2007if,LopezVal:2012zb,Broggio:2014mna}
and in particular have a large impact on the prediction of $M_W$. 
%
Because of the potentially large quantum effects, the THDM has become popular to 
provide an explanation for the larger value of the $W$ mass given by the new CDF result; 
accordingly, a series of recent papers deal with analyses of electroweak precision data 
in various versions of the THDM~\cite{Lu:2022bgw,Bahl:2022xzi,Babu:2022pdn,Heo:2022dey,Ahn:2022xeq,Han:2022juu,Arcadi:2022dmt,Ghorbani:2022vtv,Abouabid:2022lpg,Lee:2022gyf,Benbrik:2022dja,Botella:2022rte,Frandsen:2022xsz}.

The one-loop corrections in the THDM for scenarios with large mass splittings, as 
required for $M_W$, are quite sizeable; hence calculations beyond one loop are needed 
in order to get reliable theoretical predictions for precision observables.  
Since the dominating quantum effects are associated with the $\rho$-parameter, the leading
two-loop effects can be embedded via $\Delta\rho$ in case of broken custodial symmetry.
In \cite{Hessenberger:2016atw}, we presented the two-loop corrections to the $\rho$-parameter
in the $CP$-conserving THDM originating from the top-Yukawa and the scalar self interactions.
This was done under the assumption that one of the neutral $CP$-even states of the THDM can be 
identified with the scalar boson observed at the LHC with properties like the Higgs particle of the SM.

In this paper we provide the currently most precise predictions for $M_W$ 
and the leptonic $\sin^2\theta_{\rm eff}$ in the $CP$-conserving THDM 
combining the two-loop corrections of~\cite{Hessenberger:2016atw} with the 
THDM one-loop corrections and the complete set of available SM loop contributions.   
The outline is as follows: 
Section~\ref{Sec:THDM}  specifies the $CP$-conserving THDM and gives a list of
theoretical constraints on the model parameters. A phenomenologically interesting 
special case is the Inert Higgs Doublet Model (IHDM), which is briefly desribed as well. 
Section~\ref{Sec:MWMZrel}  reviews the status of the $W$-boson mass prediction in the SM 
and explains the incorporation of the available non-standard corrections for the
$W$-mass prediction in the THDM. Analogously, Section~\ref{Sec:sinthetaeff}
deals with the effective mixing angle.
Numerical results are presented in Section~\ref{Sec:Results}, with conclusions 
given in Section~\ref{Sec:Conclusions}.

\section{The Two-Higgs-Doublet model}
\label{Sec:THDM}

The Higgs sector  of the THDM contains  two $SU(2)_L$ doublets
of complex scalar fields with hypercharge $Y=1$, 
\begin{eqnarray}
\Phi_1=
\left(\begin{array}{l}
	\phi_1^+ \\
	\phi_1^0 
\end{array}\right), &\quad&
\Phi_2=
\left(\begin{array}{l}
	\phi_2^+ \\
	\phi_2^0 
\end{array}\right).
\label{scalarfields}
\end{eqnarray}
Under the assumption of a discrete $Z_2$ symmetry
($\Phi_1 \to \Phi_1, \, \Phi_2 \to -\Phi_2$)
which is only softly broken by a non-diagonal mass term,
the general gauge invariant potential is given by 
(see e.g.~\cite{Gunion:2002zf})
\begin{align}
V\left(\Phi_1,\Phi_2\right) =
&m_{11}^2 \left(\Phi _1{}^{\dagger } \Phi _1\right)+m_{22}^2 \left(\Phi _2{}^{\dagger } \Phi _2\right) 
  - m_{12}^2 \left(\left(\Phi _1{}^{\dagger } \Phi _2\right)+\left(\Phi _2{}^{\dagger } \Phi _1\right)\right)\notag\\
&+\frac{1}{2} \Lambda _1 \left(\Phi _1{}^{\dagger } \Phi _1\right){}^2
  +\frac{1}{2} \Lambda _2 \left(\Phi _2{}^{\dagger } \Phi_2\right){}^2  
  +\Lambda _3 \left(\Phi _2{}^{\dagger } \Phi _2\right) \left(\Phi _1{}^{\dagger } \Phi _1\right)\notag\\
&+\Lambda _4 \left(\Phi _1{}^{\dagger } \Phi _2\right) \left(\Phi _2{}^{\dagger } \Phi _1\right) 
   +\frac{1}{2} \Lambda _5 \left(\left(\Phi _1{}^{\dagger } \Phi _2\right){}^2
   +\left(\Phi _2{}^{\dagger } \Phi_1\right){}^2\right).
\label{Eq:potential}
\end{align}
Assuming conserved $CP$ symmetry in the Higgs sector,
all  parameters in the potential are chosen to be real. 
The minimum of the potential corresponds to the vacuum field configurations
\begin{equation}
 \langle\Phi_i\rangle=
   \frac{1}{\sqrt{2}} 
\begin{pmatrix}
  0 \\  v_i 
 \end{pmatrix} , 
\end{equation}
and the minimum conditions for $v_1$ and $v_2$ read as follows,
\begin{align}
m_{11}^2=
&m_{12}^2 \, t_{\beta }-\frac{1}{2} v^2 c_{\beta }^2\left(\Lambda _1+\Lambda _{345} t_{\beta }^2\right), \nonumber\\
m_{22}^2=
&\frac{m_{12}^2}{t_{\beta }}-\frac{1}{2} v^2 s_{\beta }^2 \left(\Lambda _2+\frac{\Lambda _{345}}{t_{\beta }^2}\right) ,
\end{align}
with
\begin{equation}
v^2=v_1^2+v_2^2 , \qquad
t_\beta \equiv \tan\beta=\frac{v_2}{v_1} , \qquad
\Lambda_{345}=\Lambda_3+\Lambda_4+\Lambda_5 .
\label{Eq:TanBeta}
\end{equation}
Expanding the scalar fields around the vacuum expectation values,
\begin{eqnarray}
\Phi_1=
\left(\begin{array}{c}
	\phi_1^+ \\
	\frac{1}{\sqrt{2}}\left(v_1+\eta_1+i\chi_1\right) 
\end{array}\right), 
\quad
 \Phi_2 =
\left(\begin{array}{c}
	\phi_2^+\\
	\frac{1}{\sqrt{2}}\left(v_2+\eta_2+i\chi_2\right)
\end{array}\right) ,
\label{Eq:scalarparam}
\end{eqnarray}
yields the mass terms quadratic in the fields, 
which are diagonalized by unitary transformations
\begin{equation}
\begin{pmatrix}
	G^\pm \\
	H^\pm 
\end{pmatrix}
=
\Rot{\beta}
\begin{pmatrix}
\phi_1^\pm \\
\phi_2^\pm \\
\end{pmatrix},\quad
\begin{pmatrix}
	G^0 \\
	A^0 
\end{pmatrix}
=
\Rot{\beta}
\begin{pmatrix}
\chi_1 \\
\chi_2 \\
\end{pmatrix},\quad
\begin{pmatrix}
H^0 \\
h^0
\end{pmatrix}
=
\Rot{\alpha}
\begin{pmatrix}
\eta_1 \\
\eta_2
\end{pmatrix},
\label{Eq:fieldrot}
\end{equation}
with
\begin{equation}
 \Rot{x}=
 \begin{pmatrix}
\cos x & \sin x \\
-\sin x & \cos x \\ 
\end{pmatrix} 
\, \equiv \,
 \begin{pmatrix}
 c_x & s_x \\
-s_x & c_x \\ 
\end{pmatrix}  ,
\end{equation}
in order to disentangle the Goldstone bosons $G^0$ and $G^\pm$ from the 
physical mass eigenstates: two $CP$-even states $h^0$ and $H^0$,
one $CP$-odd state $A^0$, and a pair $H^\pm$ of charged Higgs bosons.

The gauge-boson masses are determined by $v$ and the 
electroweak gauge couplings $g_1, g_2$,
\begin{equation}
M_W^2=\frac{1}{4} g_2^2 v^2 , \quad 
M_Z^2=\frac{1}{4} \big( g_1^2+g_2^2\big) v^2 \, ,
\label{Eq:GaugeMasses}
\end{equation}
and provide the doublet-specific relation to the  
 electroweak mixing angle,
\begin{align}
 \cos^2\theta_W&=c_W^2=\frac{M_W^2}{M_Z^2},\nonumber\\
 \sin^2\theta_W&=s_W^2=1-c_W^2.
\end{align}
%
Whereas the combination $v^2=v_1^2+v_2^2$  is fixed by the
vector boson masses, there are seven  
other free parameters of the Higgs potential, which  
can be expressed in terms of the Higgs boson masses, 
the mixing angles $\alpha$ and $\beta$,
and one remaining independent  parameter
(see~\cite{Gunion:2002zf} for the general relations). 
Conventionally the soft $Z_2$-breaking mass parameter $m_{12}^2$ is chosen. 
In our previous calculation of the two-loop contributions 
to the $\rho$-parameter~\cite{Hessenberger:2016atw}, 
we used the dimensionless quantity
\begin{equation}
\lfive=\frac{2\,m_{12}^2}{v^2\,\CB\SB}
\label{Eq:Lambda5}
\end{equation}  
from the parameterisation given in \cite{gunion:1990} instead, which
is employed in the \texttt{FeynArts} model-file of the THDM. For
consistency we present the results of the $W$ boson mass and the effective 
leptonic mixing angle also using the combination $\lfive$ in~\eqref{Eq:Lambda5}. 

Furthermore, two additional assumptions have been made in~\cite{Hessenberger:2016atw}, 
which are adopted also for the present study.
First the $CP$-even state $h^0$ is identified with the
scalar particle observed by the LHC experiments.
Second, the alignment limit \cite{Bernon:2015qea} is applied where
the couplings of $h^0$ to the gauge bosons and fermions are
identical to the corresponding couplings of the SM Higgs boson.
Formally this can be achieved by setting
\begin{equation}
 \alpha=\beta-\frac{\pi}{2}.
\end{equation} 
Flavour-changing neutral currents (FCNC) through neutral Higgs exchange at the 
tree level can be avoided by the arrangement that not more than one of the 
doublets couples to fermions of a given charge~\cite{Glashow:1976nt,Paschos:1976ay}, 
This has lead to four different model classes in the literature 
which go by the names type-I, type-II, type-X and type-Y 
(see for example the review \cite{Branco:2011iw} for more details). 
In all of them, however, the couplings between up-type quarks and
Higgs bosons are the same. 
In our calculation of the non-standard two-loop corrections in the THDM 
all the Yukawa couplings except those of the  top quark are neglected, 
hence the results of \cite{Hessenberger:2016atw} are valid independent
of the model classification.  In the alignment limit, the coupling 
between the SM-like scalar~$h^0$ and the top quark is identical 
to the top-Yukawa coupling in the SM,  while the couplings 
between the top quark and the non-standard Higgs bosons 
$A^0$, $H^0$ and $H^\pm$ are modified by an additional factor of $\TB^{-1}$.

\subsection{The Inert-Higgs-Doublet model}
\label{Sec:IHDM}

A particular interesting version of the THDM is the
Inert-Higgs-Doublet-Model (IHDM) \cite{Deshpande:1977rw}
which is distinguished by an unbroken $Z_2$ symmetry. 
Under this symmetry all the SM fields and  the doublet $\Phi_1$ 
are even, whereas $\Phi_2$ is odd.
The IHDM has received attention in the context  of radiative
neutrino masses \cite{Ma:2006km} or as a solution to the naturalness 
problem \cite{Barbieri:2006dq}. Moreover, since the $Z_2$ symmetry
is unbroken, the lightest particle that is oddly charged can provide a 
dark matter candidate \cite{LopezHonorez:2006gr}. 
An overview with more details on  phenomenology and additional references 
can be found in \cite{Belyaev:2016lok}.

The potential of the IHDM is given by~\eqref{Eq:potential} setting
$m_{12}^2=0$.  
The scalar doublets appear in the following form,
\begin{equation}
\label{Eq:IHDMphi}
 \Phi_1=
 \begin{pmatrix}
  G^\pm \\
  \frac{1}{\sqrt{2}}\left(v+h^0+i G^0\right)
 \end{pmatrix},
 \quad
 \Phi_2=
 \begin{pmatrix}
  H^\pm \\
  \frac{1}{\sqrt{2}}\left(H^0+i A^0\right)
 \end{pmatrix} .
\end{equation}
Only $\Phi_1$ has a non-vanishing vacuum expectation value $v$,
related to the parameters of~\eqref{Eq:potential} via the minimum condition,
\begin{equation}
 v^2=-\frac{2\, m_{11}^2}{\Lambda_1}.
\end{equation}
With $h^0 = H_{\text{SM}}$ and the Goldstone bosons $G^0,G^\pm$, the
doublet $\Phi_1$ is SM-like.
The second doublet $\Phi_2$, the inert doublet, contains instead the non-standard 
Higgs fields $H^0$, $A^0$ and $H^\pm$. 
The four physical masses are related to the parameters of the potential
in the following way, 
\begin{align}
 \MhO^2&= -2 m_{11}^2 = \Lambda_1 v^2 , \nonumber \\
 \MHH^2&=\musq +\frac{1}{2}\left(\Lambda_3+\Lambda_4+\Lambda_5\right)v^2,\nonumber \\
 \MAO^2&=\musq +\frac{1}{2}\left(\Lambda_3+\Lambda_4-\Lambda_5\right)v^2, \nonumber\\
 \MHp^2&=\musq +\frac{1}{2}\Lambda_3 v^2.
\end{align}
The couplings of the scalar $h^0$ to fermions and gauge bosons are
identical to the corresponding couplings of the Higgs boson in the SM. 
Owing to the $Z_2$ symmetry, the additional particles $H^0$, $A^0$ and
$H^\pm$ do not couple to fermions. Moreover, they can appear only
pairwise in interaction vertices and the lightest of these
non-standard particles will be stable. 
If this is one of the neutral bosons, the IHDM provides a suitable dark matter candidate.

In order to specify the free parameters of the IHDM, 
it is convenient to choose, besides $\MhO = M_{H_{\rm SM}}$,
the masses $\MHH$, $\MAO$, $\MHp$, 
the quartic coupling $\Lambda_2$ of the non-standard scalars, 
and the quantity
\begin{equation}
 \LIHDM=\Lambda_3+\Lambda_4+\Lambda_5,
\label{Eq:LamIHDM}
\end{equation}
which is of special interest since  $\LIHDM$ describes the coupling
of the standard-like Higgs particle to a potential dark matter candidate. 


\subsection{Theoretical constraints}
\label{Sec:Constraints}

The parameters of the potential given in \eqref{Eq:potential} are subject to
various restrictions. A stable vacuum requires the potential to be bounded 
from below. In the THDM this requirement has to be fulfilled for all possible 
directions along which the component fields of $\Phi_{1,2}$ go to large values.
As explained in~\cite{Deshpande:1977rw,Klimenko:1984qx,Maniatis:2006fs} 
the conditions
\begin{align}
 \Lambda_1&>0, \nonumber \\
 \Lambda_2&>0, \nonumber \\
 \Lambda_3+\sqrt{\Lambda_1\Lambda_2}&>0 ,\nonumber \\
 \Lambda_3+\Lambda_4-|\Lambda_5|&>-\sqrt{\Lambda_1\Lambda_2},
\end{align}
ensure that the  quartic terms in the potential are positive for large values 
of the field components in all directions. 
These tree-level bounds can be improved by considering higher-order
corrections to the potential.
For more details see the discussion in \cite{Branco:2011iw} and
references therein. For our analysis we employ the tree-level bounds 
as an estimate for the allowed parameter range.

The unitarity requirement for the scattering matrix
puts additional constraints on the parameters of the Higgs potential.
Due to the optical theorem, the $s$-wave scattering length $a_0$ is restricted 
to $|a_0|\leq1/2$. For scattering processes with four scalars in the high-energy 
limit, $a_0$ is directly proportional to the scalar couplings.
Moreover, due to the Goldstone boson equivalence theorem the scattering
of longitudinal gauge bosons can be calculated as scalar--scalar
scattering by replacing the gauge boson with the corresponding
Goldstone bosons.  In the SM, the constraints from tree-level
unitarity gives an upper bound on the Higgs mass
\cite{Lee:1977eg,Lee:1977yc}. The application of the analysis in the
THDM
\cite{Casalbuoni:1986hy,Casalbuoni:1987cz,Maalampi:1991fb,Kanemura:1993hm,Akeroyd:2000wc,Horejsi:2005da,Ginzburg:2005dt}
is more complicated due to the larger number of possible scattering
processes and the involved structure of the scalar quartic
couplings. With the help of a unitary tranformation, the scattering
matrix of the coupled scalar--scalar channels can be simplified by
using the original fields $\phi^+_i$, $\eta_i$  and $\chi_i$ from
\eqref{Eq:scalarparam} instead of the mass eigenstates. 
The restrictions on the $s$-wave scattering length constrain the eigenvalues 
of the scattering matrix at tree-level (see for example \cite{Branco:2011iw})
\begin{align}
e_{1,2}=&\frac{3}{2} \left(\Lambda _1+\Lambda _2\right)\pm\sqrt{\frac{9}{4} \left(\Lambda _1-\Lambda _2\right){}^2+\left(2 \Lambda _3+\Lambda _4\right){}^2},\nonumber\\
e_{3,4}=&\frac{1}{2} \left(\Lambda _1+\Lambda _2\right)\pm\frac{1}{2} \sqrt{\left(\Lambda _1-\Lambda _2\right){}^2+4 \Lambda _4^2},\nonumber\\
e_{5,6}=&\frac{1}{2} \left(\Lambda _1+\Lambda _2\right)\pm\frac{1}{2} \sqrt{\left(\Lambda _1-\Lambda _2\right){}^2+4 \Lambda _5^2},\nonumber\\
e_7=&\Lambda _3+2 \Lambda _4-3 \Lambda _5,\nonumber\\
e_8=&\Lambda _3-\Lambda _5,\nonumber\\
e_9=&\Lambda _3+2 \Lambda _4+3 \Lambda _5,\nonumber\\
e_{10}=&\Lambda _3+\Lambda _5,\nonumber\\
e_{11}=&\Lambda _3+\Lambda _4,\nonumber\\
e_{12}=&\Lambda _3-\Lambda _4,
\end{align}
to fulfill $|e_i|\leq 8\pi$. We are employing these tree-level bounds 
as an estimate of the validity of perturbativity. 
For more accurate restrictions higher-order corrections have  to be 
considered for the scattering processes. 
A one-loop analysis of the unitarity bounds can be found in \cite{Grinstein:2015rtl,Cacchio:2016qyh}.

The constraints from vacuum stability and $S$-matrix unitarity 
are identical in the aligned THDM and the IHDM. 
In the IHDM the parameters are further constrained by the condition
\begin{equation}
\frac{m_{11}^2}{\sqrt{\Lambda_1}}\leq\frac{\musq}{\sqrt{\Lambda_2}}
\end{equation}
to ensure that $v$ in \eqref{Eq:IHDMphi} corresponds to 
the global minimum of the potential \cite{Ginzburg:2010wa}.

\section{The {\boldmath{$M_W$-$M_Z$}} interdependence}
\label{Sec:MWMZrel}

The correlation  between the gauge boson masses $M_W$ and $M_Z$
can be established via the Fermi constant $G_F$, 
which is determined with high accuracy from precise measurements of 
the muon lifetime and  
the calculation of the muon decay width within the low-energy effective
Fermi model including QED corrections up to ${\cal O}(\alpha^2)$ for the point-like 
interactions~\cite{Behrends:1955mb,Berman:1958ti,Kinoshita:1958ru,vanRitbergen:1998yd,vanRitbergen:1999fi,Steinhauser:1999bx}.
Comparison of the muon-decay amplitude calculated in electroweak theories 
like the SM or THDM with the Fermi model result yields the relation
\begin{equation}
\frac{G_F}{\sqrt{2}} = \frac{\pi\alpha}{2 M_W^2\big(1-\frac{M_W^2}{M_Z^2} \big) }
                                  \left(1+\Delta r\right) ,
\label{Eq:MWMZrel}
\end{equation}
where the non-QED loop corrections are summarized in the quantity $\Delta r$.
Since it depends on all the virtual particles in the loop contributions,
\begin{equation}
 \Delta r=\Delta r \left(M_W,M_Z,m_t,\dots\right)
\label{Eq:deltar}
\end{equation}
is a model-dependent quantity, and the relation~\eqref{Eq:MWMZrel} 
provides the prediction of $M_W$ in specific models 
in terms of the model parameters and the highly accurate 
input quantitites $M_Z, G_F$ and the electromagnetic 
fine-structure constant $\alpha$. \\

\subsection{One-loop calculations}

At the one-loop level, the contributions to $\Delta r$ 
consist of the $W$ self-energy, vertex and box diagrams, and the
related counterterms. 
In the on-shell renormalization scheme, the required counterterms
arise from charge and mass renormalization
 (see for example \cite{Denner:1991kt,Hollik:1995dv}),
\begin{align}
e_{0}=& Z_e e =(1+\dZe{1})e,  \nonumber \\
M^2_{W,0} & =M_W^2+\dMWsq[1], \nonumber \\
M^2_{Z,0} & =M_Z^2+\dMZsq[1], 
\end{align}
according to
\begin{align}
\delta Z_e& = \frac{1}{2}\PiAA{0}+\frac{s_W}{c_W}\frac{\Self[\gamma Z]{(1)}{0}}{M_Z^2}, \nonumber\\
\dMWsq[1]&=\Re\Self[W]{(1)}{M_W^2},  \nonumber\\
\dMZsq[1]&=\Re\Self[Z]{(1)}{M_Z^2}, 
\end{align}
with the transverse part $\Sigma_{V}$ of the gauge-boson self-energies, 
the photon vacuum polarization
\begin{equation}
 \PiAA{k^2}=\frac{\Self[\gamma]{(1)}{k^2}}{k^2} ,
 \label{Eq:Vacpol}
\end{equation}
and the non-diagonal photon--$Z$ mixing self-energy. 
In  the on-shell scheme the electroweak mixing angle is a derived quantity, 
\begin{equation}
 s_{W,0}^2=1-c_{W,0}^2 \,= \, 1-\frac{M_{W,0}^2}{M_{Z,0}^2} \, = \,  
    s_W^2 +  \dSWsq + \cdots
\label{Eq:SWbare}
\end{equation}
yielding 
\begin{equation}
s_W^2  = 1 - \frac{M_W^2}{M_Z^2} , \qquad
c_W^2 = 1 - s_W^2 \, ,
\label{Eq:swdef}
\end{equation}
and the one-loop counterterm 
\begin{equation}
\frac{\dSWsq}{s_W^2}=-\frac{c_W^2}{s_W^2}\frac{\dCWsq}{c_W^2}=\frac{c_W^2}{s_W^2}\left(\frac{\dMZsq}{M_Z^2}-\frac{\dMWsq}{M_W^2}\right) .
\label{Eq:dswoneloop}
\end{equation}
With the notation $\delta_\text{vertex+box}$ for the vertex and box diagram contributions
including external wave-function renormalization, one can write
\begin{align}
\label{Eq:Deltar1}
\Delr{(1)}{}=&2\,\dZe{1}+\frac{\Self[W]{(1)}{0}-\dMWsq}{M_W^2}-\frac{c_W^2}{s_W^2}\left(\frac{\dMZsq}{M_Z^2}-\frac{\dMWsq}{M_W^2}\right)+\delta_\text{vertex+box} \, .
\end{align}
At the one-loop level, $\Delta r$ can be split 
\begin{equation}
 \Delr{(1)}{}=\Delta\alpha-\frac{c_W^2}{s_W^2}\Delrho[]{1}+\Delr{(1)}{rem}
 \label{Eq:drSM1Loop}
\end{equation}
into three parts, containing
\begin{itemize}
\item 
the shift of the fine-structure constant $\Delta\alpha$ from charge renormalization, originating 
from the light-fermion contribution to the photon vacuum polarization; 
\item
the loop correction $\Delrho[]{1}$ to the $\rho$-parameter, which and can be written as 
\begin{equation}
\label{Eq:delro1Loop}
 \Delrho{1}=\frac{\Self[Z]{(1)}{0}}{M_Z^2}-\frac{\Self[W]{(1)}{0}}{M_W^2}  \, ;
\end{equation}
\item 
the remainder part $\Delr{(1)}{rem}$ summarizing
all the other terms.
\end{itemize}
The loop correction to the $\rho$-parameter is sensitive to the mass splitting between the partners
within an isospin doublet~\cite{Veltman:1977kh}.
In the SM, this yields a sizeable contribution from the top--bottom quark 
doublet~\cite{Veltman:1977kh,Chanowitz:1978mv,Chanowitz:1978uj}. 
In the THDM, it can moreover get large contributions from the non-standard Higgs bosons
in case of mass splittings between neutral and charged scalars, 
yielding the dominant non-standard loop corrections in $\Delta r$.

In the alignment limit of the THDM, one can identify  the loop contributions
of the scalars $h^0$, $G^0$, $G^\pm$ 
with the standard scalar contributions to  the SM part of $\Delta r$.
Consequently, the  non-standard contributions to $\Delta r$ arise 
from the scalars $H^0$, $A^0$, and $H^\pm$. 
Thus, one can write at the  one-loop level
\begin{align}
\label{Eq:oneloopseparation}
 \Delr{(1)}{}   =  \Delr{(1)}{SM} +  \Delr{(1)}{NS} \, .
\end{align}
Since the scalar contributions to the
vertex and box corrections are negligible 
due to the small Yukawa couplings to electron and muon,
the non-standard scalars contribute to $\Delta r$ only
through the gauge boson self-energies, yielding
the non-standard part
\begin{align}
 \Delr{(1)}{NS}=\PiAA[\text{NS}]{0}
  +\frac{\Self[W,\text{NS}]{(1)}{0}-\dMVsq{1}{W,\text{NS}}}{M_W^2}
  -\frac{c_W^2}{s_W^2}\left(\frac{\dMVsq{1}{Z,\text{NS}}}{M_Z^2}-\frac{\dMVsq{1}{W,\text{NS}}}{M_W^2}\right).
\label{Eq:drNS1Loop}
\end{align}
The subindex NS indicates that only the non-standard parts 
of the self-energies and counterterms, respectively, have to be taken.
The corresponding expressions are listed in the Appendix. 
The dominant effect on $\Delta r$, as noted above,  can be traced back 
to the additional non-standard correction to the $\rho$-parameter,
entering \eqref{Eq:drSM1Loop} with 
\begin{align}
\Delrho[NS]{1}=\frac{\alpha}{16 \pi  s_W^2 M_W^2}\Bigg\{&  
\frac{m_{A^0}^2 m_{H^0}^2 }{m_{A^0}^2-m_{H^0}^2}\log \left(\frac{m_{A^0}^2}{m_{H^0}^2}\right)-\frac{m_{A^0}^2 m_{H^{\pm }}^2 }{m_{A^0}^2-m_{H^{\pm}}^2}\log \left(\frac{m_{A^0}^2}{m_{H^{\pm }}^2}\right)\notag\\
 &-\frac{m_{H^0}^2 m_{H^{\pm }}^2 }{m_{H^0}^2-m_{H^{\pm }}^2}\log
 \left(\frac{m_{H^0}^2}{m_{H^{\pm }}^2}\right)+m_{H^{\pm }}^2\Bigg\} .
\label{Eq:droNS1Loop}
\end{align}
The origin of $\Delrho[NS]{1}$ are the couplings of the non-standard Higgs
sector when they violate the custodial symmetry. 
$\Delrho[NS]{1}$ vanishes for
$m_{H^0} = m_{H^\pm}$ or $m_{A^0} = m_{H^\pm}$.

\subsection{Higher oder corrections in the SM}
\label{Sec:HOSM}

For the products of the large one-loop contributions $\Delta\alpha$
and $\Delrho{1}$ resummations were derived which 
incorporate $(\Dalpha)^n$ to all orders~\cite{Marciano:1979yg,Sirlin:1983ys} 
and two-loop terms of the form $(\Delta\rho^{(1)})^2$ and
$(\Dalpha\Delta\rho^{(1)})$~\cite{Consoli:1989fg}; 
moreover, prescriptions are given 
in~\cite{Consoli:1989fg} for 
incorporating also the higher-order reducible terms from
$\Dalpha$ and $\Delta\rho$.

The electroweak corrections in the SM are known at the complete two-loop 
level~\cite{Freitas:2000gg,Freitas:2002ja,Awramik:2003ee,Awramik:2002wn,Onishchenko:2002ve,Awramik:2002vu};
in addition, the pure fermion-loop corrections~\cite{Weiglein:1998jz,Stremplat:1998} 
up to four-loop order are available, as well as
the three-loop corrections to the $\rho$ parameter~\cite{Faisst:2003px}.
QCD corrections to $\Delta r$ are calculated at the two- and three-loop level,
$\order{\alpha\alpha_s}$
\cite{Djouadi:1987,Djouadi:1988,Kniehl:1990,Halzen:1991,Kniehl:1992,Djouadi:1993ss}
and $\order{\alpha\alpha_s^2}$
\cite{Avdeev:1994db,Chetyrkin:1995ix,Chetyrkin:1995js,Chetyrkin:1996cf},
and to the $\rho$ parameter at four-loop order $\order{\alpha_s^3 G_F m_t^2}$
\cite{Schroder:2005db,Chetyrkin:2006bj,Boughezal:2006xk}. 

For practical calculations,
in \cite{Awramik:2003rn} a simple parametrisation is given which
reproduces the SM prediction for $M_W$ 
from \eqref{Eq:MWMZrel} with
\begin{equation}
\Delta r=\Delta r^{\left(\alpha\right)}
+\Delta r^{\left(\alpha^2\right)}
+\Delta r^{\left(\alpha\alpha_s\right)}
+\Delta r^{\left(\alpha\alpha_s^2\right)}
+\Delta r^{\left(G_F^3 m_t^6\right)}
+\Delta r^{\left(G_F^2 m_t^4 \alpha_s\right)} 
+\Delta r^{\left(G_F m_t^2 \alpha_s^3\right)} 
\label{Eq:drcont}
\end{equation}
including $\Delta r^{\left(\alpha\right)}$ as the SM part of the one-loop result~\eqref{Eq:drSM1Loop}, 
the full two-loop electroweak correction $\Delta r^{\left(\alpha^2\right)}$,
the two- and three-loop  QCD corrections 
$\Delta r^{\left(\alpha\alpha_s\right)}$ and $\Delta r^{\left(\alpha\alpha_s^2\right)}$;
the electroweak three-loop term $\Delta r^{\left(G_F^3 m_t^6\right)}$ 
and the three- and four-loop mixed electroweak--QCD contributions
$\Delta r^{\left(G_F^2 m_t^4\alpha_s\right)}$ and $\Delta r^{\left( G_F m_t\alpha_s^3\right)}$ 
are approximations based on the corresponding corrections to the $\rho$ parameter.

\subsection{Two-loop corrections in the THDM}
\label{Sec:TwoLoopConts}

 The THDM prediction for $\Delta r$ beyond the one-loop level can be decomposed
according to
\begin{equation}
 \Delta r= \Delta r_\text{SM}+\Delta r_\text{NS},
\end{equation}
where $\Delta r _\text{SM}$ contains all the known SM corrections 
mentioned above and
$\Delta r_\text{NS}$ comprises the additional non-standard
contribution, which in the alignment case originates from the
non-standard bosons $H^0$, $A^0$ and $H^\pm$. 
The expansion up to two-loop order,
\begin{equation}
 \Delta r_\text{NS}=\Delr{(1)}{NS}+\Delr{(2)}{NS} \, ,
 \label{Eq:drNSloopexp}
\end{equation}
accommodates the complete non-standard one-loop part $\Delr{(1)}{NS}$ 
from \eqref{Eq:drNS1Loop}
and the  two-loop part $\Delr{(2)}{NS}$ which we approximate by 
including the potentially large terms associated with the $\rho$-parameter.
These contain products of   $\Delrho[t]{1}, \Delrho[NS]{1}, \, \Delta\alpha$ 
as well as 
the non-standard two-loop irreducible corrections to the $\rho$-parameter
calculated in \cite{Hessenberger:2016atw}. 
Technically, they are obtained in the gauge-less limit
(vanishing gauge couplings $g_1,g_2$ while keeping the ratio
$M_W/M_Z$ constant) and in the top-Yukawa approximation 
where only the top-quark mass is  kept different from zero.
This yields a significant step of improvement at the two-loop level taking into account
those contributions that can become sizeable whenever  the 
one-loop contribution in $\Delr{}{}$ is large.

\medskip
In analogy to the SM, the non-standard two-loop contribution can be
written in terms of a one-particle irreducible part $\Delr{(2)}{NS,irr}$
and a reducible part $\Delr{(2)}{NS,red}$ , 
\begin{equation}
 \Delr{(2)}{NS}=\Delr{(2)}{NS,red}+\Delr{(2)}{NS,irr}.
 \label{Eq:drNS2Lsplit}
\end{equation}
The reducible non-standard contribution, $\Delr{(2)}{NS,red}$, arises from 
the non-standard one-loop correction to the $\rho$-parameter~\eqref{Eq:droNS1Loop}
in the expansion of $\Delr{}{}$ up to two-loop order, in the on-shell scheme 
given by
\begin{equation}
 \Delta r= \Dalpha+\Dalpha^2-\frac{c_W^2}{s_W^2}\Delrho{1}
 \left(1+2\Dalpha-2\frac{c_W^2}{s_W^2}\Delrho{1}\right)+\dots
 \label{Eq:Delr_red}
\end{equation}
with
\begin{equation}
 \Delrho{1}=\Delrho[t]{1}+\Delrho[NS]{1} \, , \quad
 \Delrho[t]{1} = \frac{3\, \alpha}{16 \pi s_W^2}\frac{m_t^2}{ M_W^2} \, .
 \label{Eq:Delrho1Lsplit}
\end{equation}
The two-loop terms of $\Delr{}{}$ which contain only $\Dalpha$ and $\Delrho[t]{1}$
are already included in the SM part; 
therefore, the non-standard part remains as follows, 
\begin{equation}
 \Delr{(2)}{NS,red}=-2\frac{c_W^2}{s_W^2}\, \Dalpha\, \Delrho[NS]{1}
                      +4\frac{c_W^4}{s_W^4}\, \Delrho[NS]{1}\, \Delrho[t]{1}
                      +2\frac{c_W^4}{s_W^4} \left(\Delrho[NS]{1}\right)^2 \, .
 \label{Eq:delrhoNSred}
\end{equation}
As a side note, we mention that a reparametrization of  
$\Delrho[]{1} \to \Delta\overline{\rho}^{(1)} $ 
replacing the basic on-shell parameters in the normalization
of  $\Delrho[]{1} $ by the Fermi constant  
with the help of  
\begin{equation}
 \frac{G_F}{\sqrt{2}}=\frac{\pi \alpha}{2 s_W^2 M_W^2}\left(1+\Delta r^{(1)}+\dots\right)
\label{Eq:GFoneloop}
\end{equation}
induces additional factorized two-loop terms in $\Delr{}{}$
that allow to rewrite~\eqref{Eq:Delr_red} as follows,
\begin{equation}
  \Delta r= \Dalpha+\Dalpha^2- \frac{c_W^2}{s_W^2} \DelrhoG{1}
   \left(1+\Dalpha-\frac{c_W^2}{s_W^2}\, \DelrhoG{1}\right)+\dots
\end{equation}
which corresponds to the expansion of the resummed form
\begin{equation}
 1 + \Delr{}{}  \, \to \, 
 \frac{1}{1-\Dalpha} \cdot \frac{1}{1+\frac{c_W^2}{s_W^2} \Delta\overline{\rho}} 
\end{equation}
found in~\cite{Consoli:1989fg} for the SM.

\medskip
The irreducible non-standard two-loop contribution $\Delr{(2)}{NS,irr}$ 
originates from the $W$ self-energy  
and the counterterms for the parameters in~\eqref{Eq:MWMZrel}.
In our approximation, the two-loop charge renormalization constant is zero 
and the two-loop self-energy $\Self[W]{(2)}{0}$  is canceled by the two-loop
$W$ mass counterterm.
Hence, the only remaining quantity is the two-loop counterterm for $s_W^2$,
obtained from expanding the bare relation~\eqref{Eq:SWbare}, 
\begin{equation}
 \frac{\dSWsq[2]}{c_W^2} \, = \, 
   -\, \frac{\dMZsq[1]}{M_Z^2} \left(\frac{\dMZsq[1]}{M_Z^2}-\frac{\dMWsq[1]}{M_W^2} \right)
   \, +\, \frac{\dMZsq[2]}{M_Z^2}-\frac{\dMWsq[2]}{M_W^2} \, ,
  \label{Eq:SW2LoopCT}
\end{equation}
which in the gauge-less limit reduces to
\begin{equation}
 \frac{\dSWsq[2]}{c_W^2}\, =\, 
               -\, \frac{\Self[Z]{(1)}{0}}{M_Z^2}\Delrho{1} \,
              +\, \frac{\Self[Z]{(2)}{0}}{M_Z^2}-\frac{\Self[W]{(2)}{0}}{M_W^2}  
              \, = \,   \Delrho{2}  \, ,
\label{Eq:Deltarho2}
\end{equation}
where $\Delta\rho^{(2)}$
is the two-loop correction to the $\rho$ parameter as derived in~\cite{Hessenberger:2016atw}. 

The $W,Z$ self-energies  $\Sigma^{(2)}_{W,Z}$ correspond to the set of
genuine two-loop diagrams and diagrams with subloop renormalization, 
i.e.~one-loop diagrams with insertion of mass countertems and of  $\dSWsq[1]$. 
The two-loop terms with $\dSWsq[1]$ factorize and can be extracted; 
in combination with the factorized first term in~\eqref{Eq:Deltarho2}
one obtains
\begin{equation}
 \Delta\rho^{(2)}=-\frac{c_W^2}{s_W^2}\left(\Delrho{1}\right)^2+\delrho{2} \, .
\label{Eq:withreduciblepart}
\end{equation}
The residual part $\delrho{2}$ contains the genuine two-loop self-energies 
combined with the part of the subloop renormalization from the mass counterterms 
of the internal particles. 
The quadratic term in $\Delrho{1}$ from \eqref{Eq:withreduciblepart} 
is already included in $\Delr{2}{NS,red}$ 
(in the $G_F$-parametrization of $\Delta\rho$ it would  vanish).
The irreducible two-loop part of $\Delta r$ is thus given by
\begin{equation}
\Delr{(2)}{irr} \, =\, -  \frac{c_W^2}{s_W^2} \, \delrho{2} \, ,
\label{Eq:Delrirr}
\end{equation}
it still contains the SM contribution.
In the notation of~\cite{Hessenberger:2016atw}, $\delrho{2}$ is divided into four finite parts
\begin{equation}
\label{Eq:aufteilung}
 \delta\rho^{(2)} = \delta\rho^{(2)}_{\rm t, SM} + \delta\rho^{(2)}_{\rm t, NS} 
 + \delta\rho^{(2)}_{\rm H,NS} + \delta\rho^{(2)}_{\rm H, Mix}  ,
\end{equation}
where  $\delrho[t,SM]{2}$ contains the SM-like scalars $h^0$, $G^0$, $G^\pm$ 
and the top-Yukawa coupling, whereas the non-standard scalars 
$H^0$, $A^0$, $H^\pm$ appear in the residual three parts.

Separating off the SM contribution $\delrho[t,SM]{2}$ in~\eqref{Eq:Delrirr}, 
which is already part of $\Delta r_{\rm SM}$,
one obtains the non-standard contribution as follows,
\begin{align}
\Delr{(2)}{NS,irr}  & =\,  \Delr{(2)}{t,NS} + \Delr{(2)}{H,NS} + \Delr{(2)}{H,Mix} \nonumber\\
                           & = \, - \frac{c_W^2}{s_W^2} \, \big(
                                 \delrho[t,NS]{2} + \delrho[H,NS]{2} + \delrho[H,Mix]{2} \big) .
\label{Eq:DelrNSirr}
\end{align}                                   

\noindent
The various entries of $\Delta r$ have the following properties.                                    

\begin{itemize}
\item 
$\Delr{(2)}{t,NS}$ incorporates the coupling of the top quark to the non-standard scalars, 
which is proportional to $\TB^{-1}$ and enters all the diagrams for this contribution quadratically. 

\item 
$\Delr{(2)}{H,NS}$ incorporates the non-standard scalar  interactions. 
It is proportional to the squared coupling between three non-standard scalars, which has the form
 \begin{equation}
  \frac{1}{2}\left(\frac{1}{\TB}-\TB\right)\frac{\left(2\MHH^2-\lfive v^2\right)}{v}
  \label{Eq:3NSCoupling}
 \end{equation}
in the alignment limit. Consequently, $\Delr{(2)}{H,NS}$ vanishes for 
 \begin{equation}
  \TB=1 \quad
   {\rm  or}   \quad 
   \lfive=\frac{2\MHH^2}{v^2}.
 \end{equation}
 Moreover, it is zero for $\MAO=\MHp$, 
 since this mass configuration restores the custodial symmetry 
 of the scalar potential.
However,  differently from $\Delrho[NS]{1}$, 
this contribution does not vanish for $\MHH=\MHp$ (if $\MAO\neq\MHp$). 

\item 
$\Delr{(2)}{H,Mix}$ incorporates 
the interaction between the standard and the non-standard scalars. 
Similarly to $\Delrho[NS]{1}$, 
it vanishes for $\MHH=\MHp$ or $\MAO=\MHp$ due to the restoration of
the custodial symmetry. 
Differently from $\Delrho[NS]{1}$, 
this contribution contains additional couplings between  $h^0$ and the non-standard scalars of the form
 \begin{equation}
  \frac{m_{h^0}^2+2m_S^2-\lfive v^2} {v^2},\quad (S=H^0, A^0, H^\pm),
  \label{Eq:MixCoupling}
 \end{equation}
which can be enhanced by $\lfive$. 
In contrast to $\Delr{(2)}{t,NS}$ and $\Delr{(2)}{H,NS}$, 
the contribution $\Delr{(2)}{H,Mix}$ 
does not depend on $\TB$. 
\end{itemize}

\subsection{Loop corrections in the IHDM}

The non-standard one-loop corrections to the gauge-boson self-energies
are the same in the IHDM and the general THDM in the alignment limit, 
yielding $\Delr{(1)}{\text{NS}}$ for both cases which depends only on
the masses of the $H^0, A^0, H^\pm$ bosons.  
Differences arise at the two-loop level.
Since Yukawa interactions of the non-standard scalar doublet 
are absent, there is no  $\Delr{(2)}{t,NS}$ term in the IHDM. 
Moreover, as discussed in \cite{Hessenberger:2016atw}, 
the only non-standard two-loop contribution to the $\rho$ parameter, 
$\delrho[IHDM]{2}$, follows from the interaction of $h^0$, $G^0$, $G^\pm$ with 
$H^0$, $A^0$, $H^\pm$ and is thus equivalent to $\delrho[H,Mix]{2}$; 
it vanishes for equal charged and non-standard neutral Higgs-boson masses 
and contains the coupling between $h^0$ and the non-standard scalars, 
which is determined by the combination  $\LIHDM$ 
of coefficients in the scalar potential, as specified in~\eqref{Eq:LamIHDM}
of Section~\ref{Sec:IHDM}.  
The corresponding contribution to $\Delta r$ is given by
\begin{equation}
\Delr{(2)}{\text{IHDM,irr}}=\frac{c_W^2}{s_W^2}\delrho[IHDM]{2} \, ,
\label{Eq:DelrIHDMirr}
\end{equation}
which introduces an additional dependence on the IHDM parameter  $\LIHDM$, 
which is not present in  $\Delr{(1)}{NS}$ and $\Delr{(2)}{NS,red}$.

\subsection{Incorporation of the higher-order corrections}

For an accurate prediction of $M_W$ in the THDM we have 
to combine all the by now available loop contributions 
from the SM and beyond in the quantity $\Delr{}{}$, which
depends on  the SM input and on all the free non-standard 
parameters of the THDM,
\begin{equation}
  \Delta r (M_W,\dots) = \Delta r_{\rm SM}(M_W,\dots)
                       + \Delta r_{\rm NS}(M_W,\dots)   
\label{Eq:Deltarfull}
\end{equation}
The SM part, $\Delta r_{\rm SM}$, 
contains, besides the one-loop result, the complete two-loop
and the partial higher-order contributions listed in~\eqref{Eq:drcont}.
Since the two-loop electroweak part 
is quite an inconvenient expression
which involves furthermore numerical integrations,
we make use of the parametrization given in~\cite{Awramik:2006uz}
in terms of the SM input parameters.

\medskip
Summarizing the various standard and non-standard contributions to $\Delta r$ 
we write \eqref{Eq:Deltarfull} as follows,
\begin{align}
 \Delta r & \, =  \, 
        \Delta r_{\text{SM}} 
      +\Delr{(1)}{NS} 
      +\Delr{(2)}{NS,red} 
        +\Delr{(2)}{NS,irr} 
\end{align}
where the three non-standard terms are specified in the 
equations~\eqref{Eq:drNS1Loop},
\eqref{Eq:delrhoNSred}, and \eqref{Eq:DelrNSirr}.
In case of the IHDM, $\Delr{(2)}{NS,irr}$
has to replaced by $\Delr{(2)}{IHDM,irr}$ in~\eqref{Eq:DelrIHDMirr}.

\medskip
The predicted value of $M_W$ fulfills the
relation~\eqref{Eq:MWMZrel} for a given specific set of parameters.
Since $\Delta r$ itself does depend on $M_W$, the solution 
of~\eqref{Eq:MWMZrel} for $M_W$ has to be determined numerically.

\section{The effective electroweak mixing angle}
\label{Sec:sinthetaeff}

The electroweak mixing angle in the effective leptonic vertex of  the $Z$ boson 
\cite{Consoli:1989pc}
is another important precision observable, measured with high accuracy at the
$Z$ resonance by the LEP and SLC experiments~\cite{ALEPH:2005ab}.
Theoretically it is derived from the ratio of the dressed leptonic vector and axial vector 
couplings $g_{V,A}$ of the $Z$ boson,
 \begin{equation}
\sin^2\theta_\text{eff} \, \equiv\, s^2_l  \, =\, 
\frac{1}{4} \left(1- {\rm Re} \frac{g_V}{g_A}\right) ,
\label{Eq:sineffdef}
\end{equation}
keeping the notation $s^2_l$ as a short term.
The effective couplings
\begin{align}
 g_V & = v + \Delta g_V , \qquad  g_A = a + \Delta g_{\!A} ,  
\label{Eq:effcoup}
\end{align}
contain the lowest-order leptonic $Z$ couplings
\begin{align}
   v & =-\frac{1}{2}+2 s_W^2,   \qquad
    a  =  -\frac{1}{2} ,
\label{Eq:treelevelcoup}
\end{align}
and the corresponding loop contributions 
$\Delta g_{V,A}$ evaluated at the $Z$-mass scale
(without the QED corrections associated with virtual photons).

It is convenient to relate  $s^2_l$ to the on-shell quantity $s_W^2$
in (\ref{Eq:swdef}) by a factor $\kappa$ wich incorporates the
loop corrections  in terms of $\Delta\kappa$,
\begin{equation}
 s_l^2 \, =\,   s_W^2\, \kappa = s_W^2\left(1+\Delta\kappa\right).
 \label{Eq:sl-Definition}
\end{equation}
Analogously to $\Delta r$ in (\ref{Eq:deltar}),
$\Delta\kappa=\Delta\kappa(M_W,M_Z,m_t,\dots)$ 
is a model-dependent quantity, depending on all parameters of
the standard and non-standard particles which enter the loop contributions.
An additional model dependence in the prediction of $s^2_l$
occurs via $s_W^2$ through the $W$ mass obtained by means of $\Delta r$,
as described in the previous Section~\ref{Sec:MWMZrel}.

\subsection{One-loop calculations}
Loop corrections lead to deviations $\Delta\kappa$
from the relation $\kappa=1$. 
The expansion of~(\ref{Eq:sineffdef}) yields the
 one-loop contribution
\begin{equation}
\DelK{1} =\,  - \frac{1}{4s_W^2}\, \frac{v}{a}
 \left( \frac{\Delta g_V^{(1)}}{v} - \frac{\Delta g_A^{(1)}}{a} \right) ,
\end{equation}
which can be written as follows,
\begin{equation}
\DelK{1}  = \, \frac{c_W}{s_W}\, \frac{\Re\Self[\gamma Z]{(1)}{M_Z^2}}{M_Z^2}
                \,+\,  \frac{\dSWsq}{s_W^2}
                 \,+\, \frac{v}{v-a} 
                \left( \frac{\FV{}} {v}  -  \frac{\FA{}} {a} \right). 
\end{equation}
The first two terms originate from photon--$Z$ mixing and from
renormalization of the on-shell weak mixing angle~(\ref{Eq:dswoneloop})
The corrections from the vertex diagrams and the wave-function renormalization 
of the external fermions are collected in the vector and axial vector
form factors $F_{V,A}$, with lepton masses neglected.

In the alignment limit of the THDM, one can identify  the one-loop contribution
of the scalars $h^0$, $G^0$, $G^\pm$ 
with the standard scalar contribution to  the SM part of $\Delta\kappa$.
Consequently, the  non-standard contribution to $\Delta\kappa$ arise 
from the scalars $H^0$, $A^0$, and $H^\pm$. 
Thus, one can write at the  one-loop level
\begin{align}
\label{Eq:swoneloopseparation}
 \DelK[]{1} = \DelK[SM]{1} + \DelK[NS]{1} \, ,
\end{align}
with the SM part 
\begin{equation}
\DelK[SM]{1}  = \,
   \frac{c_W}{s_W}\, \frac{\Re\Self[\gamma Z,\text{SM}]{(1)}{M_Z^2}}{M_Z^2} 
   \,+\,\frac{\delta^{(1)}s_{W,\text{SM}}^2}{s_W^2}
   \, +\,\frac{v}{v-a}
    \left(
   \frac{F_{V,\text{SM}}^{}} {v}-\frac{F_{A,\text{SM}}^{}} {a} 
   \right)  ,
\label{Eq:DelKSM1Loop}
\end{equation}
where the subindex indicates that only the standard particles are kept
in the one-loop self-energy and vertex corrections.

The contributions from the non-standard scalars 
to the vertex corrections and to the lepton self-energies 
are suppressed owing to the small Yukawa couplings. 
Accordingly,  we can neglect the non-standard contributions to the
form factors $F_{V,A}$. 
The one-loop non-standard contribution to $\Delta\kappa$ is thus given by
\begin{equation}
 \DelK[NS]{1} =\,
   \frac{c_W}{s_W}\, \frac{\Re\Self[\gamma Z,\text{NS}]{(1)}{M_Z^2}}{M_Z^2}
   \,+\, \frac{\delta^{(1)}s^2_{W,\text{NS}}}{s_W^2}  \, ,
\end{equation}
and is obtained via the non-standard content of the vector-boson self-energies
from Appendix~\ref{App:selfenergies}.

A sizeable part of the one-loop contribution to $\Delta\kappa$ is associated
with $\Delta\rho^{(1)}$, Eq.~(\ref{Eq:delro1Loop}), via
\begin{equation}
 \DelK[]{1} = \frac{c^2_W}{s^2_W} \, \Delta\rho^{(1)} \, + \, \cdots
\end{equation}
involving the non-standard term~(\ref{Eq:droNS1Loop}),
which becomes large in case of a large mass splitting in the 
non-standard scalar spectrum.

\subsection{Higher order corrections in the SM}

The one-loop calculation of the effective mixing angle  
was performed in \cite{Marciano:1980pb} 
for the neutral-current vertex in neutrino scattering.
The first results for $s^2_l$ at the $Z$ resonance\cite{Consoli:1989pc,Gambino:1994} 
are based on one-loop calculations improved by higher orders from
$\Delta \alpha$ and $\Delta \rho$,
and an expansion in the top-quark mass \cite{Degrassi:1996ps}.
QCD corrections are known at the two-loop order 
\cite{Djouadi:1987,Djouadi:1988,Kniehl:1990,Halzen:1991,Kniehl:1992,Djouadi:1993ss}, 
together with the leading three-loop
\cite{Avdeev:1994db,Chetyrkin:1995ix,Chetyrkin:1995js} 
and four-loop \cite{Schroder:2005db,Chetyrkin:2006bj,Boughezal:2006xk} 
terms from the top quark. 
The two-loop electroweak SM contributions have been obtained in
\cite{Awramik:2004ge,Hollik:2005va,Hollik:2005ns,Hollik:2006ma,Awramik:2006ar,Awramik:2006uz}. The leading three-loop corrections via the $\rho$ parameter at
$\mathcal{O}\left(G_F^3m_t^3\right)$ 
and $\mathcal{O}\left(G_F^2\alpha_s^2m_t^2\right)$ 
were calculated for a massless Higgs boson in \cite{vanderBij:2000cg} 
and with Higgs mass dependence in \cite{Faisst:2003px}.

In \cite{Awramik:2006uz} a simple parametrization is given, which incorporates 
the complete electroweak one- and two-loop corrections together with the QCD 
corrections of $\order{\alpha\alpha_s}$ 
\cite{Djouadi:1987,Djouadi:1988,Kniehl:1990,Halzen:1991,Kniehl:1992,Djouadi:1993ss} 
and $\order{\alpha\alpha_s}$ 
\cite{Avdeev:1994db,Chetyrkin:1995ix,Chetyrkin:1995js} 
and the leading electroweak three-loop corrections of 
$\mathcal{O}\left(G_F^3m_t^3\right)$ and
$\mathcal{O}\left(G_F^2\alpha_s^2m_t^2\right)$ 
\cite{vanderBij:2000cg,Faisst:2003px}. 
An update is found in~\cite{Dubovyk:2019szj} which includes also the
leading $\order{G_F m_t^2\alpha_s^3}$ terms
\cite{Schroder:2005db,Chetyrkin:2006bj,Boughezal:2006xk};
the differences, however, are very small, 
not more than $2\cdot 10^{-5}$ for $s^2_l$.
The required input parameters are the masses of the Higgs particle, the top quark 
and the $Z$ boson, together with the strong coupling constant 
$\alpha_s\left(M_Z^2\right)$.

\subsection{Two-loop corrections in the THDM}

The one-loop corrections can be supplemented by the leading reducible 
and irreducible two-loop contributions, containing 
products of $\Dalpha$ and $\Delrho{1}$ and the
irreducible two-loop correction $\delrho{2}$ to $\rho$, 
see Eq.~(\ref{Eq:aufteilung}).
The resulting two-loop contribution to $\Delta\kappa$ is given by
\begin{align}
 \Delta\kappa^{(2)}
&=\Dalpha\, \frac{c_W^2}{s_W^2}\Delrho{1}
   -\frac{c_W^4}{s_W^4}\left(\Delrho{1}\right)^2
   +\frac{c_W^2}{s_W^2}\, \delrho{2}
 \label{Eq:DelKappa2L}
\end{align}
As a side-remark, we note that the appearance of the reducible products 
in $\Delta\kappa$ is again a consequence 
of the on-shell parameters in the normalization of $\Delta\rho$. 
After a reparametrization $\Delrho{1} \to \Delta \overline{\rho}^{(1)}$ 
by the Fermi constant with the help of 
Eqs.~(\ref{Eq:GFoneloop})
and (\ref{Eq:drSM1Loop})
in the expression
\begin{equation}
 \DelK{1}=\frac{c_W^2}{s_W^2}\, \Delrho{1}+\dots
\end{equation}
the resulting two-loop shift cancels the reducible term in \eqref{Eq:DelKappa2L} 
and we obtain
\begin{equation}
 \Delta\overline{\kappa}^{(2)} = \frac{c_W^2}{s_W^2}\, \delta\rho^{(2)}
\end{equation}
as the two-loop correction to $\Delta\kappa$.

Next we have to identify the non-standard two-loop corrections.
Separating  $\Delrho[t]{1}$ and $\Delrho[NS]{1}$ 
in the reducible term of (\ref{Eq:DelKappa2L}) yields
\begin{align}
 \Dalpha\, \frac{c_W^2}{s_W^2}\Delrho{1}
   -\frac{c_W^4}{s_W^4} \left(\Delta\rho^{(1)}\right)^2 \, = \;\;
 & \Dalpha\, \frac{c_W^2}{s_W^2}\left(\Delrho[t]{1}+\Delrho[NS]{1}\right)\notag\\
 & -\frac{c_W^4}{s_W^4}\left(\left(\Delrho[t]{1}\right)^2 
    +2\,\Delrho[t]{1}\Delrho[NS]{1} +\left(\Delrho[NS]{1}\right)^2\right).
\end{align}
Since the terms without $\Delrho[NS]{1}$ are already incorporated in
$\Delta\kappa_\text{SM}$,  we retain the non-standard reducible part
\begin{equation}
 \DelK[NS,red]{2}\, =\,\Dalpha\,\frac{c_W^2}{s_W^2} \Delrho[NS]{1}
  -\frac{c_W^4}{s_W^4}\left(2\, \Delrho[t]{1}\Delrho[NS]{1}
  +\left(\Delrho[NS]{1}\right)^2\right).
\label{Eq:dkapNSred}
\end{equation}

The irreducible part $\delta\rho^{(2)}$ can be separated into the different non-standard 
contributions according to \eqref{Eq:aufteilung}, which leads to the following 
contributions to $\Delta\kappa$,
\begin{itemize}
 \item contribution from the non-standard top-Yukawa couplings 
 \begin{equation}
  \DelK[t,NS]{2}=\frac{c_W^2}{s_W^2}\, \delrho[t,NS]{2},
 \end{equation}
 \item contribution from the exclusively non-standard scalar corrections
 \begin{equation}
  \DelK[H,NS]{2}=\frac{c_W^2}{s_W^2}\, \delrho[H,NS]{2},
 \end{equation}
 \item contribution from the mixed standard--non-standard scalar corrections
 \begin{equation}
 \DelK[H,Mix]{2}=\frac{c_W^2}{s_W^2}\, \delrho[H,Mix]{2}.
 \end{equation}
\end{itemize}
In the IHDM, the only irreducible  non-standard two-loop contribution 
to the $\rho$ parameter is $\delrho[IHDM]{2}$
(which is equivalent to $\delrho[H,Mix]{2}$), hence
\begin{equation}
 \DelK[IHDM,irr]{2}=\frac{c_W^2}{s_W^2}\, \delrho[IHDM]{2}.
\end{equation}
With these specifications, the two-loop irreducible contribution 
to $\Delta\kappa$ is given by
\begin{equation}
 \DelK[NS,irr]{2}=\DelK[t,NS]{2}+\DelK[H,NS]{2}+\DelK[H,Mix]{2}
\end{equation}
in the aligned THDM,  and 
\begin{equation}
 \DelK[NS,irr]{2}=\DelK[IHDM,irr]{2}
\end{equation}
in the IHDM.

\subsection{Incorporation of the higher-order corrections}

For an accurate prediction of $s^2_l$ in the THDM we have 
to combine all the by now available loop contributions 
from the SM and beyond in the quantity $\Delta\kappa$, which
depends on  the SM input quantitites and on  the non-standard 
parameters of the THDM,
\begin{equation}
  \Delta \kappa (M_W,\dots) = \Delta \kappa_{\text{SM}}(M_W,\dots)
                       + \Delta \kappa_{\text{NS}}(M_W,\dots)   
\label{Eq:Deltakappafull}
\end{equation}
For a given  set of parameters, the predicted value of $s^2_l$ 
fulfills the relation~\eqref{Eq:sl-Definition}
with the corresponding $M_W$ calculated from~\eqref{Eq:MWMZrel}. 
Since the $W$ mass varies with the model parameters, 
$\Delta\kappa$ is needed as an expression where $M_W$ appears as
an independent variable

\medskip
In the following we write only $M_W$ as an explicit variable in $\Delta \kappa$,
dropping the ellipsis. The SM part,
\begin{align}
\Delta \kappa_{\text {SM}}(M_W) & =\, \Delta \kappa ^{(1)} _{\text{ SM}} (M_W)
                                        + \Delta \kappa ^{\text{(ho)}} _{\text{SM}} (M_W)
\label{Eq:DelkappaSMfull}
\end{align}
contains besides the one-loop result the complete two-loop
and the partial higher-order contributions listed above.
Since the full SM result is an inconvenient expression,
we make use of the parametrization for the SM prediction of $s^2_l$
given in~\cite{Awramik:2006uz} for the set of SM input parameters
$M_Z, m_t, m_{h^0},\alpha_s$, yielding the value  $\slsm$.
By inverting~\eqref{Eq:sl-Definition} one obtains 
\begin{equation}
\Delta \kappa_{\text{SM}}  (\mwsm)
\, =\,  s_{l,\text{SM}}^2 \,
 \left(1-\frac{\mmwsm}{M_Z^2} \right)^{-1} \, -\, 1 \, ,
\label{Eq:dkapsef}
\end{equation}
while $M_W$ is simultaneously fixed by the standard value $\mwsm$
via \eqref{Eq:MWMZrel} with $\Delta r_{\text{SM}}$.

Since the $W$ mass varies with the non-standard model parameters, we need
$\Delta\kappa$ as an expression where $M_W$ appears as an
independent variable.
For the one-loop correction of $\Delta\kappa$ in the SM we use the explicit result 
\eqref{Eq:DelKSM1Loop} with $M_W$ as one of the input quantities.
From~(\ref{Eq:dkapsef}) we obtain
\begin{equation}
\Delta \kappa ^{\text{(ho)}} _{\text{SM}} (\mwsm)
 =\Delta\kappa_{\text{SM}}(\mwsm) -\DelK[SM]{1}\left(M_{W,\text{SM}}\right)
 \, \equiv\, \delta\kappa_\text{SM}^{\text{(ho)}}
\end{equation}
which we use as an approximation for the second term in~(\ref{Eq:Deltakappafull}),
yielding
\begin{equation}
 \Delta\kappa_\text{SM}(M_W)
=\DelK[SM]{1}\left(M_W\right) + \delta\kappa_\text{SM}^{(\text{ho})}.
\end{equation}
The prediction of the effective leptonic mixing angle 
can now be written in the following way,
\begin{equation}
s^2_l  =
 \left( 1- \frac{M_W^2}{M_Z^2} \right)
  \left( 1+\Delta\kappa_{\text{SM}} 
+\DelK[NS]{1}+\DelK[NS,red]{2}
 +\Delta\kappa^{(2)}_\text{NS,irr}  \right)
\end{equation}
evaluated for
$M_W$ as obtained from~\eqref{Eq:MWMZrel} with the full
$\Delta r = \Delta r_{\text{SM}} + \Delta r_{\text{NS}}$.

\medskip
For comparison, we will also need the effective mixing angle
including just the one-loop non-standard corrections, 
calculated from
\begin{equation}
 \sin^2\theta_\text{eff}^{(1)}  \equiv  s^{2\, (1)}_l  \, =\, 
           \left(1-\big(\frac{M_W^{(1)}}{M_Z}\big)^2\right)
           \left(1+\Delta\kappa_\text{SM}+\DelK[NS]{1}\right),
\end{equation}
where $M_W^{(1)}$ is used as the corresponding input for the $W$-boson mass 
obtained from~\eqref{Eq:MWMZrel} with 
$\Delta r = \Delta r_{\text{SM}}  + \Delta r_{\text{NS}}^{(1)}$.

\clearpage

\section{Numerical results}
\label{Sec:Results}

For the analysis of the precision observables the following input~\cite{ParticleDataGroup:2020ssz}
for the SM parameters is used, together with the mass of the SM-like Higgs boson 
$m_{h^0}=125.1\text{ GeV}$,
\begin{align}
 &  G_F =1.1663787 \cdot 10^{-5} \text{ GeV}^{-2} , \quad
     M_Z=91.1876\text{ GeV} , 
       \nonumber \\
 & m_t=172.76\text{ GeV} , \quad 
    \alpha_s\left(M_Z^2\right)=0.1179 .
\label{Eq:SMpara_Input}
\end{align}
The shift $\Dalpha=0.05907$
in the electromagnetic fine structure constant,
$\Delta \alpha=\Delta \alpha_\text{lept}+\Delta \alpha_\text{had}$,
contains the leptonic  contribution 
up to four-loop order \cite{Steinhauser:1998rq,Sturm:2013uka}
$\Delta\alpha_\text{lept}=0.031498$,
and the  hadronic part  
extracted from experimental data with the help of dispersion relations,
$\Delta\alpha_\text{had}=0.027572$
taken from \cite{Jegerlehner:2001wq}; this yields $\Dalpha$ 
as the default value in the SM 
parametrizations for $M_W$, $\Delta r$ and  $s_l^2$
given in~\cite{Awramik:2003rn} and~\cite{Awramik:2006uz}.

\subsection{Parameter dependence}

\begin{figure}[t]
\centering
\includegraphics[width=\textwidth]{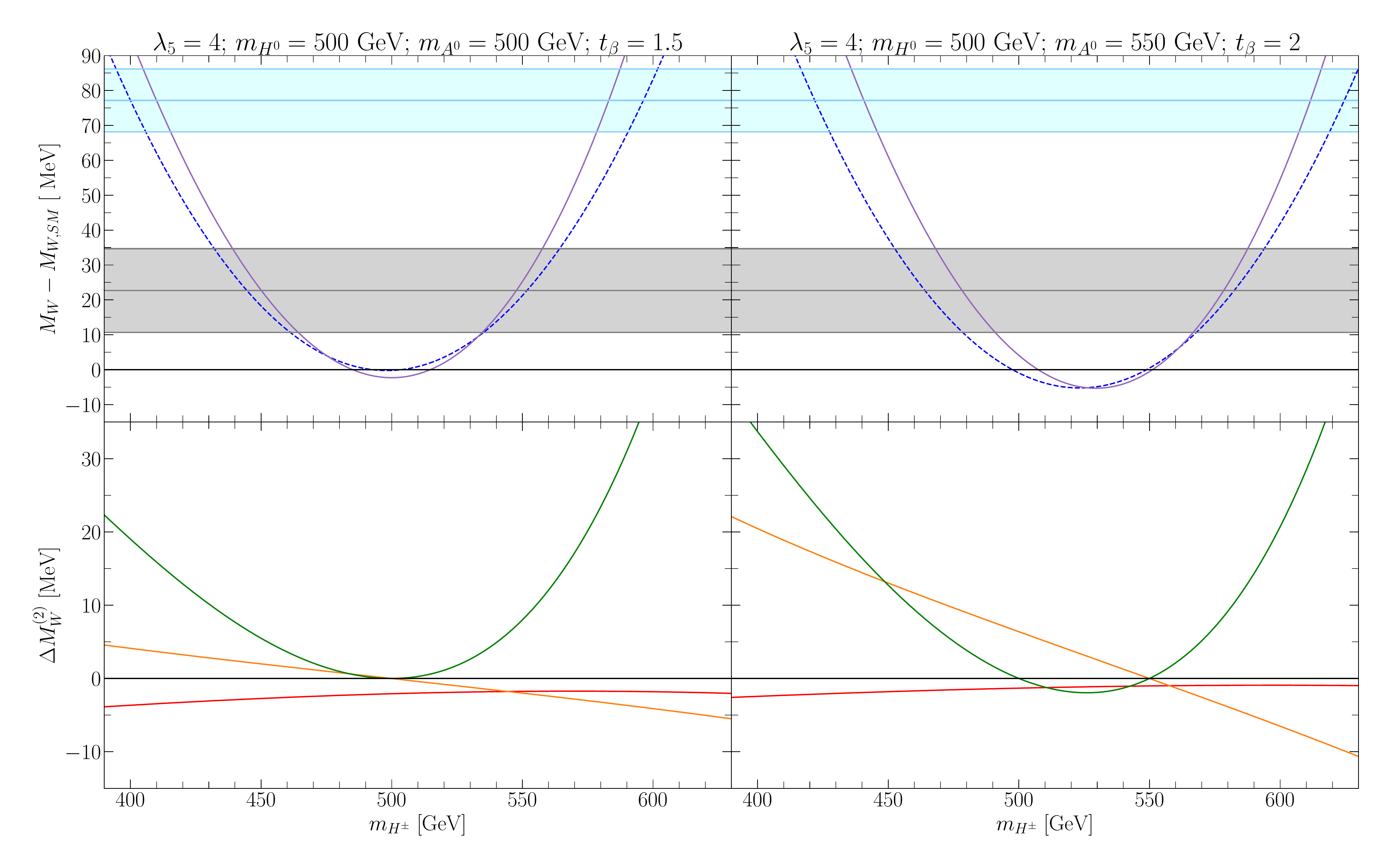}
 \caption{Shift in $M_W$ from non-standard Higgs boson mass splittings.
The upper panels show the difference to the SM result, at one-loop order (blue dashed line)
and with all the available non-standard corrections (purple line).
The current world average with $1\,\sigma$ range is displayed by the grey area;
the new CDF result is indicated by the shaded area in light blue.
The lower panels show the individual two-loop contributions from $\Delr{(2)}{t,NS}$ (red line),  
$\Delr{(2)}{H,NS}$ (orange line), and $\Delr{(2)}{H,Mix}$  (green line).}
 \label{Fig:Res_Massdiff_W}
\end{figure}

Since the non-standard loop contribution is very sensitive to the difference between the masses 
of the charged and neutral scalars, we illustrate in Fig.~\ref{Fig:Res_Massdiff_W} the impact of a variation of
$\MHp$ on $M_W$. The masses of the neutral Higgs bosons are kept fixed, assuming equal masses 
of $H^0$ and $A^0$ on the left side and a small mass difference on the right hand side.
The settings for the other non-SM parameters are specified above the corresponding plots. 
The upper panels display the deviation of $M_W$ from the SM prediction with just the 
one-loop non-standard correction included (blue dashed line) 
and including all the available non-standard corrections (purple solid line) of two-loop order.
The grey shaded area displays the experimental result with the associated  $1\,\sigma$  uncertainty
according to the PDG value in~(\ref{Eq:Wexp}); the light-blue area in the upper part of the figure  
shows the new CDF result~(\ref{Eq:WCDF}). 
The lower panels display the effects of the individual two-loop corrections, with the reducible
contributions subtracted. The different lines correspond to the various non-standard 
two-loop contributions to $\Delta r$ yielding a mass shift of $M_W$ calculated
as function of $\Delr{(2)}{x}$
according to
\begin{equation}
 \Delta M_W^{(2)}\left(\Delr{(2)}{x}\right)=M_W\left(\Delr{(1)}{NS}+\Delr{(2)}{NS,red}
+\Delr{(2)}{x}\right)-M_W\left(\Delr{(1)}{NS}+\Delr{(2)}{NS,red}\right)
 \label{Eq:DelMWdef}
\end{equation}
with $\Delr{(2)}{x}=\Delr{(2)}{t,NS}$ (red line), $\Delr{(2)}{x}=\Delr{(2)}{H,NS}$ (orange line) 
and $\Delr{(2)}{H,Mix}$ (green line). 

\medskip
The one-loop non-standard contribution $\Delr{(1)}{NS}$ 
depends only on the scalar masses and is independent $\TB$ and $\lfive$. 
It is dominated by the non-standard correction $\Delrho[NS]{1}$ to the $\rho$ parameter,
which increases quadratically with the mass difference between charged and neutral scalars and 
is small for $\MHp\simeq\MHH$ or $\MHp\simeq\MAO$. 
Adding the non-standard two-loop corrections leads to notable deviations. 
Similiar to the one-loop contribution, the mass shifts from $\Delr{(2)}{H,Mix}$
are small for $\MHH\simeq\MHp$ or $\MAO\simeq\MHp$ and grow with the mass difference 
between charged and neutral scalars. The shifts from $\Delr{(2)}{H,NS}$
are zero for $\MHp=\MAO$ but not for $\MHp=\MHH$. 
These terms are additionally enhanced for large values of $\TB$ and vanish for $\TB=1$.
The mass shifts from $\Delr{(2)}{t,NS}$ originating from the top-Yukawa contribution to $\rho$
are substantially smaller than those from the scalar sector.
They are of some influence only 
for $\TB\simeq 1$ and are suppressed for larger values of $\TB$.
Since the SM prediction of $M_W$ is below the experimental $1\,\sigma$ limits,  also for the
current PDG world average, mass splitting between charged and neutral scalars can improve 
the agreement between the measurement and the theoretical prediction, 
in which the growing significance of the two-loop corrections is clearly visible.

Whereas for the PDG value moderate non-standard corrections  
are needed, a substantial upward shift of $M_W$ is required in order
to reach the  1$\sigma$ interval for the new CDF value, 
where the two-loop corrections are important: they enhance the one-loop result  
sizeably and hence have a considerable impact on constraining the
appropriate range of the corresponding  THDM parameters.  

\begin{figure}[b]
\centering
\includegraphics[width=\textwidth]{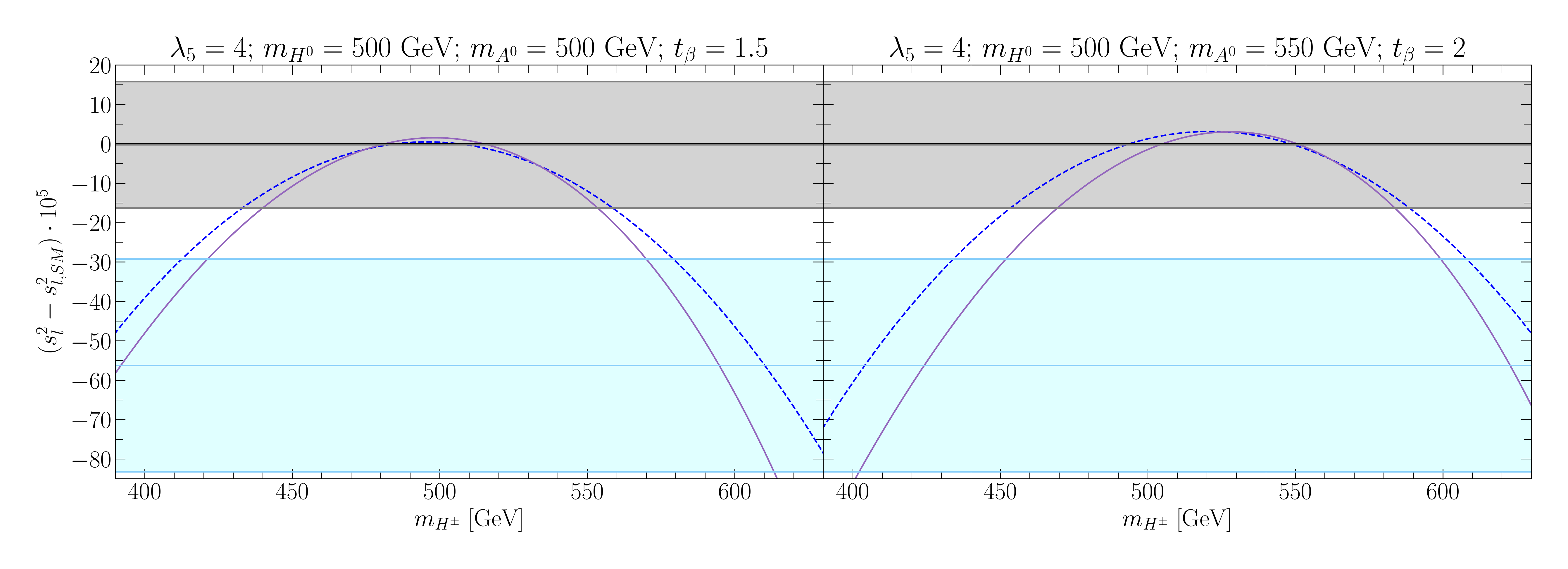}
 \caption{Shift in $s_l^2$ from non-standard Higgs boson mass splittings, for the same
input as in Fig.~\ref{Fig:Res_Massdiff_W}.
The upper panels show the difference to the SM result, at one-loop order (blue dashed line)
and with all the available non-standard corrections (purple line). The current 
world average with $1\,\sigma$ range is displayed by the upper grey area. The value 
measured by SLD is indicated by the shaded area in light blue.}  
\label{Fig:Res_Massdiff_sl}
\end{figure}

\smallskip
In a similar way,
Fig.~\ref{Fig:Res_Massdiff_sl} illustrates the effect of  a mass splitting between
charged and neutral scalars, showing the impact of a variation of $\MHp$ on the leptonic  
$\sin^2\theta_{\rm eff}  (= s^2_l)$, 
for the same parameters as in Fig.~\ref{Fig:Res_Massdiff_W} 
to allow a direct comparison of the two figures.
The masses of the neutral Higgs bosons are kept fixed, assuming equal masses of
$H^0$ and $A^0$ on the left side and a small mass difference on the right hand side.

\clearpage

\clearpage

\smallskip
Differently to the $W$ mass,  increasing the mass splitting between neutral and charged Higgs bosons
distorts the good agreement between theory and the world-average  $\sin^2\theta_{\rm eff,exp}$
in~(\ref{Eq:slep_exp}) for the effective mixing angle. The values of $\MHp$ required for getting  to the 
CDF result for~$M_W$ pull $\sin^2\theta_{\rm eff}$ away from the world average towards 
the $2\,\sigma$ boundary. It is interesting to observe that instead full  agreement 
is achieved with the SLD measurement~(\ref{Eq:slep_SLD}) from the left--right asymmetry.

\vspace*{0.5cm}

\begin{figure}[htb]
\centering
\includegraphics[width=\textwidth]{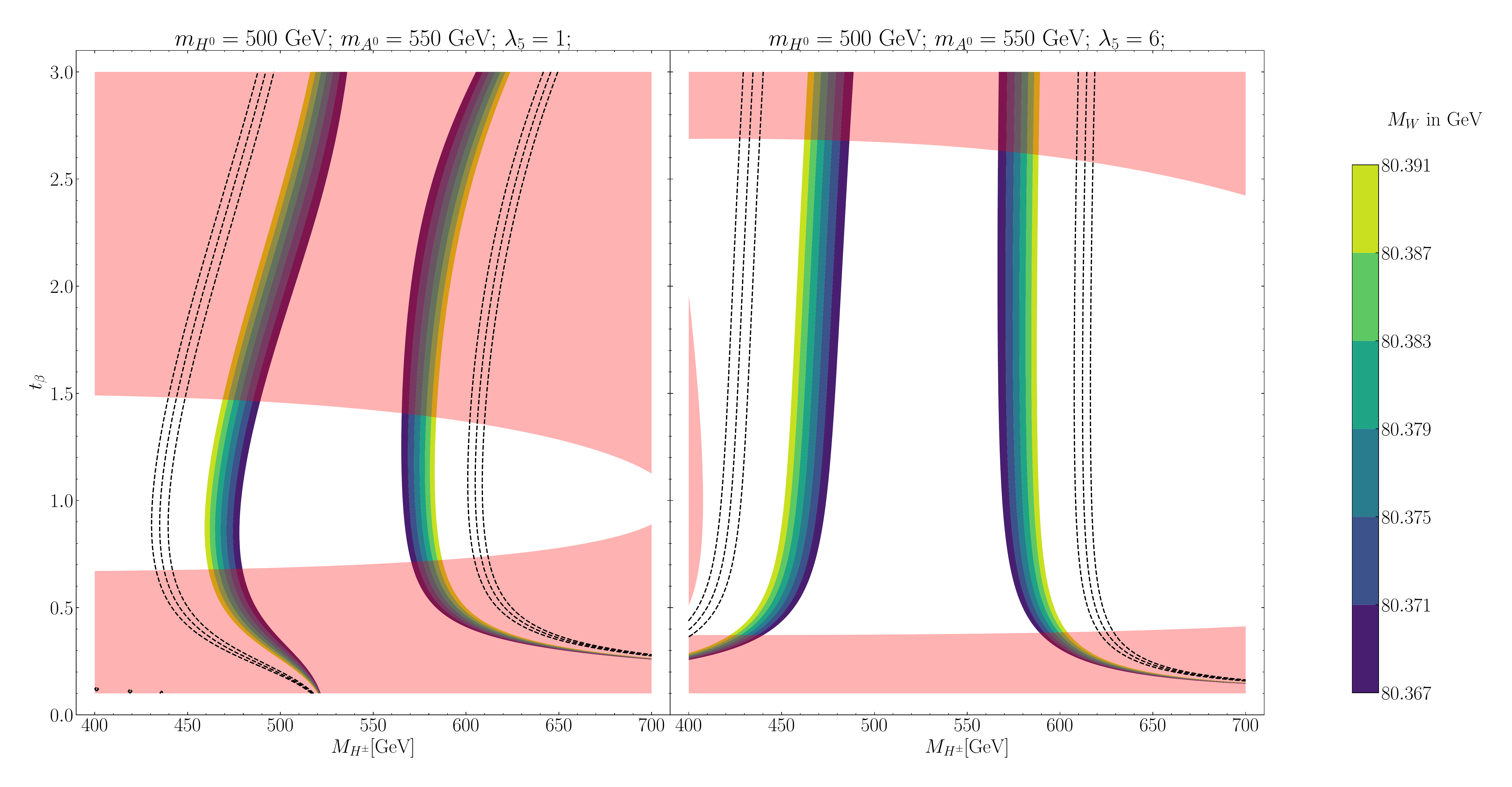}  
\caption{Influence of $\MHp$ and $\TB$ on the theoretical prediction
of $M_W$. The bands indicate the parameter configurations 
for which the calculated $M_W$ is 
within the $1\sigma$ uncertainties of the measured values,
the PDG value refers to the coloured band and the CDF value
to the dashed band. 
The red-shaded areas show the parameter regions that are excluded 
by vacuum stability and tree-level unitarity. }
\label{Fig:MHp-TB-Contours}
\end{figure}

In order to illustrate the influence of the additional parameters
$t_\beta$ and $\lambda_5$, we display in Fig.~\ref{Fig:MHp-TB-Contours}  
regions in the $\MHp$--$\TB$ plane where the predicted $M_W$ is in
agreement with the experimental results, either with the PDG value 
in~(\ref{Eq:Wexp}) or with the new CDF result in~(\ref{Eq:WCDF}).
We present results for $\lfive=1$ (left side) and $\lfive=6$ (right side) 
with neutral masses $\MHH=500\text{ GeV}$ and $\MAO=550\text{ GeV}$.
The parameter configurations
that lead to a prediction in accordance with the $1\sigma$ range
of the experimental values are illustrated by the coloured band (PDG)
and by the dashed band (CDF). 
Moreover, the areas shaded in red display the parameters which are excluded
by the theoretical bounds from vacuum stability and tree-level unitarity. 
Around $\TB=1$ the contribution via $\delrho[H,NS]{2}$ is close to zero 
and the main two-loop effects arise from $\delrho[H,Mix]{2}$, which are
additionally enhanced for smaller values of $\lfive$.
Larger $\TB$ values yield additional contributions fom $\delrho[H,NS]{2}$,
especially for $\MHp$ close to $\MHH$
when both $\Delrho[NS]{1}$ and $\delrho[H,Mix]{2}$ are small.
For $\lfive=1$ the theoretical bounds restrict the allowed values of $\TB$ 
quite strongly; nevertheless, the contributions from $\delrho[H,NS]{2}$ 
can be sizeable, giving rise to the curvature of the bands. 
For $\lfive=6$ the theoretical constraints give room to  a wider range of $\TB$, 
simultaneously the influence of $\delrho[H,NS]{2}$ is weaker, 
owing to the factor~(\ref{Eq:3NSCoupling}),
such that the dependence on $\TB$ becomes rather flat for larger values.
For very small values of $\TB$, on the other hand, 
the top-Yukawa contribution $\delrho[t,NS]{2}$ 
can become important due to its enhancement by $\TB^{-2}$. 
However, such small values are strongly restricted from flavour physics 
(see for example \cite{Enomoto:2015wbn}).

\subsection{Inert-Higgs-Doublet Model}

The predictions of $M_W$ and $s_l^2$ in the IHDM 
are dominantly influenced by mass differences  
between the neutral and charged Higgs bosons as well. 
When only the one-loop corrections $\Delr{(1)}{NS}$ and
$\DelK[NS]{1}$ are considered, the results are equivalent to those
in the aligned THDM, depending only on the non-standard particle 
masses.
Also the reducible corrections from products of $\Delrho[NS]{1}$ and
$\Dalpha$ will be identical. 
Special features occur at the two-loop level.
Differences arise through the irreducible
corrections: contributions corresponding to
$\delrho[t,NS]{2}$ and $\delrho[H,NS]{2}$ are absent in the IHDM, and
the corrections result from the remaining contribution $\delrho[IHDM]{2}$,
which corresponds to $\delrho[H,Mix]{2}$. This term contains
both standard and non-standard scalars with couplings involving
the parameter combination $\LIHDM$,  Eq.~(\ref{Eq:LamIHDM}).
Hence, the two-loop corrections induce a dependence 
of the precision observables on $\LIHDM$ as the only model parameter 
in addition to the masses entering the predictions.

\begin{figure}[htb]
\centering
 \includegraphics[width=1.05\textwidth]{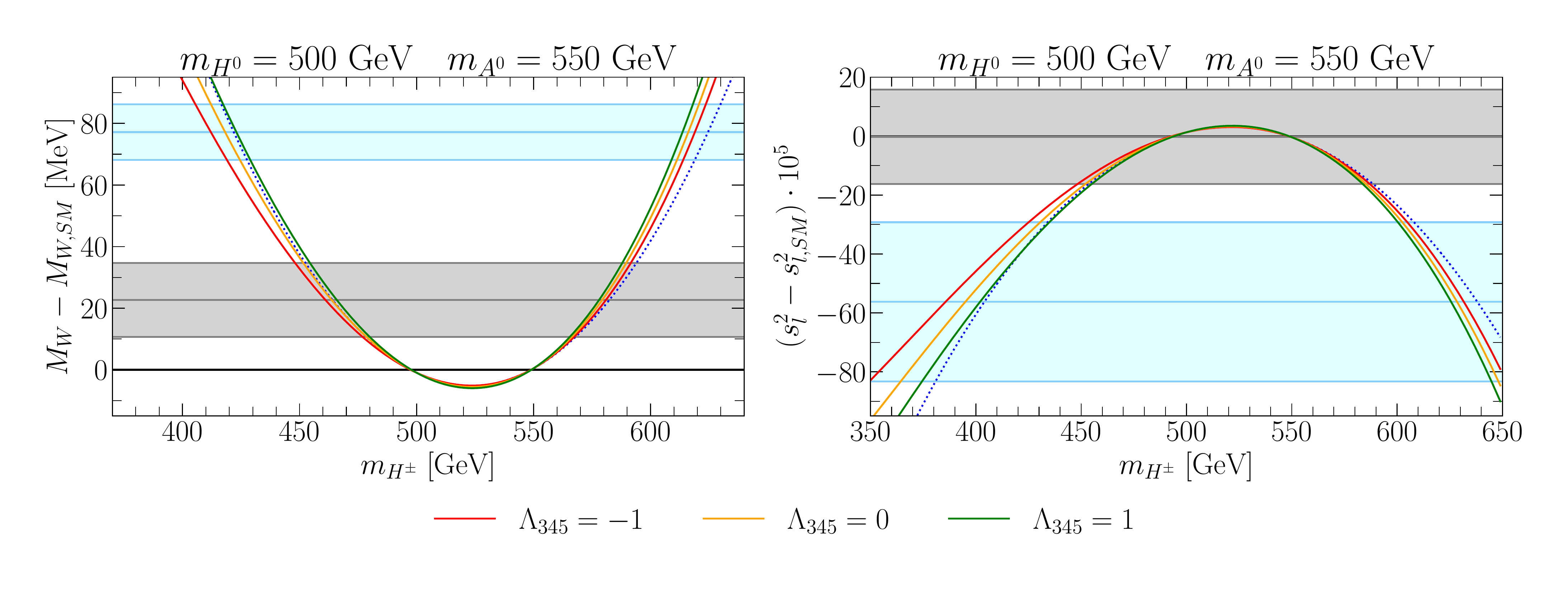}
 \caption{Shift in  $M_W$ and $s_l^2$ in the IHDM from neutral--charged 
Higgs boson mass splitting, with one-loop non-standard corrections only
(blue dashed line) and including the non-standard two-loop corrections  (full lines).
The different solid lines represent different values of $\LIHDM$. 
The current world averages with $1\sigma$ range are displayed by the grey areas;
the areas in light blue indicate the new CDF result for $M_W$ (left panel)
and the SLD result for $s^2_l$ (right panel).}
 \label{Fig:MW-MHp_IHDM}
\end{figure}

\begin{figure}[htb]
\centering
 \includegraphics[width=0.6\textwidth]{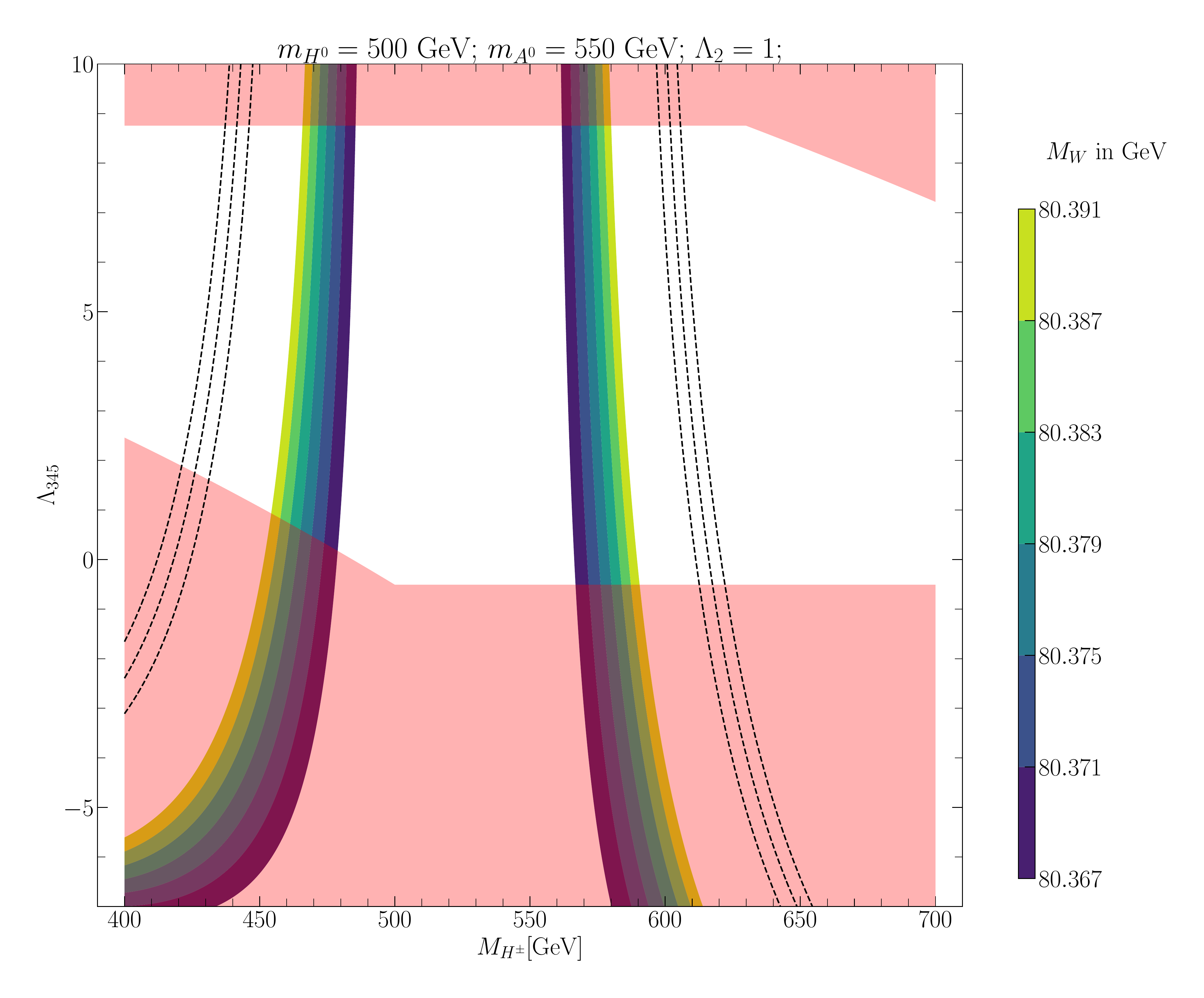}
\caption{Results in the IHDM in the $\MHp$--$\LIHDM$ plane. 
The bands indicate the parameter configurations 
for which the calculated $M_W$ is 
within the $1\sigma$ uncertainties of the measured values;
the PDG value refers to the coloured band and the CDF value
to the dashed band. 
The red-shaded areas show the parameter regions that are excluded 
by vacuum stability and tree-level unitarity. }
\label{Fig:MHpL345Contour}
\end{figure}

In Fig.~\ref{Fig:MW-MHp_IHDM} we exemplify the dependence of $M_W$ and $s_l^2$ 
on the mass difference between charged and neutral Higgs bosons for  different 
values of $\LIHDM$, in terms of their shifts with respect to the SM predictions.
In order to emphasize the effects of mass splitting
we allow also the case with $H^\pm$ as the lightest scalar, 
although in such a scenario the IHDM does not provide a dark matter candidate.
For comparison, the  shifts in $M_W^{(1)}$ and $\sin^2\theta^{(1)}_\text{eff}$, 
calculated with one-loop non-standard corrections only,
are displayed by the  blue dashed line. As expected,
the two-loop corrections show a significant dependence on the value
of $\LIHDM$ which can enhance or diminish the shifts in an asymmetric way
with respect to the minimum.

\smallskip
In Fig.~\ref{Fig:MHpL345Contour} we visualize values in the 
$\MHp$--$\LIHDM$ plane that lead to an agreement between the 
theoretical prediction and the experimental $1\sigma$ limits 
for $M_W$, again for the PDG result (coulored band) and 
the recent CDF result (dashed band). 
The areas shaded in red  
display the theoretical constraints on the  parameter space. 
Note that, differently to the two-loop corrections to the $\rho$ parameter, 
the theoretical constraints are affected also by the choice of 
the coefficient~$\Lambda_2$ in the scalar potential. 
For the selected value $\Lambda_2=1$ the bounds are rather loose, and
the allowed regions are most significantly constrained via the $W$ mass.
The growing influence of $\LIHDM$ with the mass splitting is visible
by the curvature of the dashed band.

\subsection{Computational tools}

For computation of the electroweak precision observables the code
\texttt{THDM\_EWPOS} has been
developed, which contains all the entries described in this paper and in [88].
For the development, the calculation of the one-loop corrections was done with the help of the 
Mathematica packages \texttt{FeynArts}~\cite{Hahn:2000kx} and 
\texttt{FormCalc}~\cite{Hahn:1998yk}.
The non-standard two-loop contributions were calculated using the
Mathematica package \texttt{TwoCalc}~\cite{Weiglein:1993hd,Weiglein:1992rj}.
The one- and two-loop results have been implemented as Fortran routines with the help of 
\texttt{FormCalc} and the methods from \cite{Hahn:2015gaa}, which were developed for the 
implementation of the two-loop Higgs-mass corrections at $\order{\alpha_t^2}$ in \texttt{FeynHiggs}.
These routines are then called in \texttt{THDM\_EWPOS} in the calculation of the 
various precision observables.
For the numerical evaluation of the one-loop integrals the program uses the library 
\texttt{LoopTools} \cite{Hahn:1998yk}.
For the numerical evaluation of the required two-loop integrals the Fortran routine implemented  
in the program \texttt{FeynHiggs}~\cite{Heinemeyer:1998yj,Hahn:2009zz,Bahl:2018qog}
 is used.

\section{Conclusions}
\label{Sec:Conclusions}

We have given an overview over the currently most precise calculation of the
electroweak precision observables $M_W$ and $\sin^2\theta_{\text{eff}}$
in the aligned Two-Higgs-Doublet Model
where one of the $CP$-even scalars ($h^0$) is identified with the scalar boson
at 125 GeV observed by the LHC experiments ATLAS and CMS.
It includes, besides all the known SM loop corrections, the full non-standard
one loop contributions and the leading non-standard two-loop contributions
from the top-Yukawa coupling and the self-couplings of the Higgs bosons.
As already at the one-loop level, the non-standard two-loop contributions 
from the scalar self-interactions are  particularly sensitive to mass splittings 
between neutral and charged scalars and become sizeable 
whenever the one-loop corrections are large.
Hence, they turn out to be of special importance when a large 
loop-induced shift in the predicted $W$-boson mass is required, 
as it is the case for the recently published final result of the CDF experiment.
Simultaneously, the predictions for $\sin^2\theta_{\text{eff}}$ move away
from the experimental world average by more than one standard deviation,
towards the 2~$\sigma$ limit, but show full agreement with the individual
value of the SLD experiment measured via the left--right asymmetry.

\smallskip
As a new feature, the two-loop contributions have a significant dependence on the 
parameters $\tan\beta$ and $\lambda_5$  (or $m_{12}^2$, respectively),
a coefficient of the THDM scalar potential that is not fixed by the masses 
of the neutral and charged Higgs bosons.\footnote{This was already observed
in~\cite{LopezVal:2012zb} where higher-order terms were incorporated
by means of effective couplings in the one-loop Higgs contributions.}
In the Inert-Higgs-Doublet Model,  the two-loop contributions 
depend on a specific combination $\LIHDM$ of quartic coefficients in
the scalar potential and are independent of $\tan\beta$. In both cases,
they can modify the one-loop  predictions substantially and are relevant
for phenomenological analyses.

 \smallskip
For computation of the electroweak precision observables the Fortran code
\texttt{THDM\_EWPOS} has been developed and is available at~\cite{THDMEWPOS}.
It contains all the entries described in this paper and in [88].

\newpage

\begin{appendix}

\section{Non-standard contributions to the gauge boson self-energies}
\label{App:selfenergies}

The part of  the gauge-boson self-energies in the aligned THDM resulting from the 
non-standard scalars in the one-loop diagrams is given by the following expressions,
\begin{align}
 \Self[W,\text{NS}]{(1)}{p^2} =\;
& \frac{\alpha}{16 \pi  s_W^2}  
    \left[2 A_0(m_{H^\pm}^2)+A_0(m_{H^0}^2)+A_0(m_{A^0}^2)\right.\notag\\
&\left.-4 B_{00}(p^2,m_{A^0}^2,m_{H^\pm}^2)-4 B_{00}(p^2,m_{H^0}^2,\MHp^2)\right] , \\[0.3cm]
 \Self[Z,\text{NS}]{(1)}{p^2} =\;
& \frac{\alpha }{16 \pi \, c_W^2 s_W^2} 
   \left[2 \left(c_W^2-s_W^2\right)^2 A_0(\MHp^2)+A_0(m_{H^0}^2)+A_0(m_{A^0}^2)\right.\notag\\
&\left.-4 (\left(c_W^2-s_W^2\right){}^2 B_{00}(p^2,\MHp^2,\MHp^2)
     +B_{00}(p^2,m_{A^0}^2,m_{H^0}^2))\right], \\[0.3cm]
 \Self[\gamma,\text{NS}]{(1)}{p^2} = \;
&\frac{\alpha }{2 \pi } \left[A_0(m_{H^\pm}^2)-2 B_{00}(p^2,m_{H^\pm}^2,m_{H^\pm}^2)\right] ,
\label{Eq:Sgg}
\\[0.3cm]
 \Self[\gamma Z,\text{NS}]{(1)}{p^2} = \;
&\frac{\alpha}{4 \pi} \, \frac{c_W^2-s_W^2} {c_W s_W}  
   \big[ A_0(m_{H^\pm}^2)-2 B_{00}(p^2,m_{H^\pm}^2,m_{H^\pm}^2) \big] .
\label{Eq:SgZ}
\end{align}
They are specified in terms of the one- and two-point integrals $A_0$ and $B_0$
in dimensional regularization with dimension $D$ and scale parameter $\mu$,  
 \begin{align}
 A_0(m^2) & =\, \frac{(2\pi \mu)^{4-D}} {i\pi^2}
      \int \! d^Dq \; \frac{1} {q^2-m^2+i\epsilon}  \notag \\[0.1cm]
 & =\, m^2 \left( \frac{2}{4-D} -\gamma +\log 4\pi 
   \,-\, \log \frac{m^2}{\mu^2} +1 \right)  + {\cal O}(4-D)\, , \notag \\[0.3cm]
B_0(p^2,m_1^2,m_2^2)  & =\, \frac{(2\pi \mu)^{4-D}} {i\pi^2}
   \int\! d^Dq \;
    \frac{1}{ \big[ q^2-m_1^2+i\epsilon \big] 
                   \big[ (p+q)^2-m_2^2+i\epsilon \big] }  \notag \\[0.2cm]
 & =\, \frac{2}{4-D} -\gamma +\log 4\pi 
- \int_0^1 \! dx \, 
        \log \frac{x^2 p^2 -x(p^2+m_1^2-m_2^2) - i\epsilon}{\mu^2} 
        + {\cal O}(4-D) \, ,\notag 
\notag
\end{align}
together with the tensor integral coefficients
\begin{align}
B_1(p^2,m_1^2,m_2^2) & =\, \frac{1}{2p^2} 
\left[A_0(m_1^2)-A_0(m_2^2)-(p^2+m_1^2-m_2^2)\,
                       B_0(p^2,m_1^2,m_2^2)\right] , \notag \\[0.1cm]
B_{00}(p^2,m_1^2,m_2^2)  & =\, \frac{1}{2(D-1)}
\left[A_0(m_2^2)+2m_1^2\, B_0(p^2,m_1^2,m_2^2)
    +(p^2+m_1^2-m_2^2)\, B_1(p^2,m_1^2,m_2^2) \right] . \notag
\end{align}
A compact analytic expression for $B_0$ can be found in \cite{Denner:1991kt}.

\medskip 
The relation
\begin{equation}
 B_{00}(0,m^2,m^2)=\frac{1}{2}A_0(m^2)  \notag
\end{equation}
ensures that the non-standard contributions to the photon self-energy~(\ref{Eq:Sgg}) 
and to the photon--$Z$ mixing~(\ref{Eq:SgZ})  vanish for $p^2 = 0$.

\end{appendix}

\newpage

\bibliographystyle{h-physref4}
\bibliography{MWPaper}

\begin{thebibliography}{100}

\bibitem{Aad:2012tfa}
G.~Aad {\em et~al.} [ATLAS],
\newblock Phys.Lett. {\bf B716}, 1 (2012), arXiv:1207.7214.

\bibitem{Chatrchyan:2012ufa}
S.~Chatrchyan {\em et~al.}  [CMS],
\newblock Phys.Lett. {\bf B716}, 30 (2012), arXiv:1207.7235.

\bibitem{Schael:2013ita}
S.~Schael {\em et~al.}
[ALEPH, DELPHI, L3, OPAL  and LEP Electroweak Working Group], 
\newblock Phys. Rept. {\bf 532}, 119 (2013), arXiv:1302.3415.

\bibitem{Aaltonen:2013iut}
T.~Aaltonen {\em et~al.} [CDF, D0],
\newblock Phys. Rev. {\bf D88}, 052018 (2013), arXiv:1307.7627.

\bibitem{ATLAS:2017rzl}
M.~Aaboud \textit{et al.} [ATLAS],
Eur. Phys. J. C \textbf{78},  110 (2018)
[erratum: Eur. Phys. J. C \textbf{78},  898 (2018)],
arXiv:1701.07240



\bibitem{ParticleDataGroup:2020ssz}
P.~A.~Zyla \textit{et al.} [Particle Data Group],
PTEP \textbf{2020},  083C01 (2020)


\bibitem{CDF:2022hxs}
T.~Aaltonen \textit{et al.} [CDF],
Science \textbf{376}, 170 (2022)



\bibitem{ALEPH:2005ab}
S.~Schael {\em et~al.}
[ALEPH, DELPHI, L3, OPAL, SLD, LEP Electroweak Working Group, 
SLD Electroweak Group and SLD Heavy Flavour Group],
\newblock Phys. Rept. {\bf 427}, 257 (2006), arXiv:hep-ex/0509008.



\bibitem{SLD:2000leq}
K.~Abe \textit{et al.} [SLD],
Phys. Rev. Lett. \textbf{84}, 5945 (2000),
arXiv:hep-ex/0004026 


\bibitem{Sirlin:1980}
A.~Sirlin,
\newblock Phys. Rev. D {\bf 22}, 971 (1980).


\bibitem{Marciano:1980pb}
W.~J.~Marciano and A.~Sirlin,
Phys. Rev. D \textbf{22}, 2695 (1980)
[erratum: Phys. Rev. D \textbf{31}, 213 (1985)]


\bibitem{Djouadi:1987}
A.~Djouadi and C.~Verzegnassi,
\newblock Physics Letters B {\bf 195}, 265  (1987).

\bibitem{Djouadi:1988}
A.~Djouadi,
\newblock Il Nuovo Cimento A {\bf 100}, 357 (1988).


\bibitem{Consoli:1989fg}
M.~Consoli, W.~Hollik, and F.~Jegerlehner,
\newblock Phys. Lett. {\bf B227}, 167 (1989).


\bibitem{Kniehl:1990}
B.~A. Kniehl,
\newblock Nucl. Phys. {\bf B347}, 86  (1990).

\bibitem{Halzen:1991}
F.~Halzen and B.~A. Kniehl,
\newblock Nucl. Phys. {\bf B353}, 567  (1991).



\bibitem{Kniehl:1992}
B.~A. Kniehl and A.~Sirlin,
\newblock Nucl. Phys. {\bf B371}, 141  (1992).


\bibitem{Djouadi:1993ss}
A.~Djouadi and P.~Gambino,
\newblock Phys.Rev. {\bf D49}, 3499 (1994), arXiv:hep-ph/9309298.


\bibitem{Freitas:2000gg}
A.~Freitas, W.~Hollik, W.~Walter and G.~Weiglein,
Phys. Lett. B \textbf{495}, 338 (2000), 
[erratum: Phys. Lett. B \textbf{570}, 265 (2003)],
arXiv:hep-ph/0007091



\bibitem{Freitas:2002ja}
A.~Freitas, W.~Hollik, W.~Walter and G.~Weiglein,
Nucl. Phys. B \textbf{632}, 189 (2002),
[erratum: Nucl. Phys. B \textbf{666}, 305 (2003)],
arXiv:hep-ph/0202131 

\bibitem{Awramik:2003ee}
M.~Awramik and M.~Czakon,
\newblock Phys.Lett. {\bf B568}, 48 (2003), arXiv:hep-ph/0305248.

\bibitem{Awramik:2002wn}
M.~Awramik and M.~Czakon,
\newblock Phys.Rev.Lett. {\bf 89}, 241801 (2002), arXiv:hep-ph/0208113.


\bibitem{Onishchenko:2002ve}
A.~Onishchenko and O.~Veretin,
\newblock Phys.Lett. {\bf B551}, 111 (2003), 
arXiv:hep-ph/0209010.

\bibitem{Awramik:2002vu}
M.~Awramik, M.~Czakon, A.~Onishchenko, and O.~Veretin,
\newblock Phys.Rev. {\bf D68}, 053004 (2003), arXiv:hep-ph/0209084.


\bibitem{Degrassi:2014sxa}
G.~Degrassi, P.~Gambino and P.~P.~Giardino,
JHEP \textbf{05}, 154 (2015), 
arXiv:1411.7040

\bibitem{Avdeev:1994db}
L.~Avdeev, J.~Fleischer, S.~Mikhailov, and O.~Tarasov,
\newblock Phys.Lett. {\bf B336}, 560 (1994), arXiv:hep-ph/9406363.

\bibitem{Chetyrkin:1995ix}
K.~G. Chetyrkin, J.~H. K{\"u}hn, and M.~Steinhauser,
\newblock Phys.Lett. {\bf B351}, 331 (1995), arXiv:hep-ph/9502291.

\bibitem{Chetyrkin:1995js}
K.~G. Chetyrkin, J.~H. K{\"u}hn, and M.~Steinhauser,
\newblock Phys.Rev.Lett. {\bf 75}, 3394 (1995), arXiv:hep-ph/9504413.

\bibitem{Chetyrkin:1996cf}
K.~G. Chetyrkin, J.~H. K{\"u}hn, and M.~Steinhauser,
\newblock Nucl. Phys. {\bf B482}, 213 (1996), arXiv:hep-ph/9606230.

\bibitem{Faisst:2003px}
M.~Faisst, J.~H. K{\"u}hn, T.~Seidensticker, and O.~Veretin,
\newblock Nucl. Phys. {\bf B665}, 649 (2003), arXiv:hep-ph/0302275.

\bibitem{vanderBij:2000cg}
J.~J. van~der Bij, K.~G. Chetyrkin, M.~Faisst, G.~Jikia, and T.~Seidensticker,
\newblock Phys. Lett. {\bf B498}, 156 (2001), arXiv:hep-ph/0011373.


\bibitem{Schroder:2005db}
Y.~Schr{\"o}der and M.~Steinhauser,
\newblock Phys. Lett. {\bf B622}, 124 (2005), arXiv:hep-ph/0504055.

\bibitem{Chetyrkin:2006bj}
K.~G. Chetyrkin, M.~Faisst, J.~H. K{\"u}hn, P.~Maierhofer, and C.~Sturm,
\newblock Phys. Rev. Lett. {\bf 97}, 102003 (2006), arXiv:hep-ph/0605201.

\bibitem{Boughezal:2006xk}
R.~Boughezal and M.~Czakon,
\newblock Nucl. Phys. {\bf B755}, 221 (2006), arXiv:hep-ph/0606232.


\bibitem{Consoli:1989pc}
M.~Consoli, W.~Hollik and F.~Jegerlehner,
Electroweak Radiative Corrections for $Z$ Physics,
in: {\em Z Physics at LEP~1}, 
CERN Yellow Reports \textbf{89-08} (1989),


\bibitem{Gambino:1994}
P.~Gambino and A.~Sirlin,
Phys. Rev. D \textbf{49}, 1160 (1994),
arXiv:hep-ph/9309326


\bibitem{Degrassi:1996ps}
G.~Degrassi, P.~Gambino and A.~Sirlin,
Phys. Lett. B \textbf{394}, 188 (1997),
arXiv:hep-ph/9611363



\bibitem{Awramik:2004ge}
M.~Awramik, M.~Czakon, A.~Freitas and G.~Weiglein,
Phys. Rev. Lett. \textbf{93}, 201805 (2004), 
arXiv:hep-ph/0407317


\bibitem{Hollik:2005va}
W.~Hollik, U.~Meier, and S.~Uccirati,
\newblock Nucl. Phys. {\bf B731}, 213 (2005), arXiv:hep-ph/0507158.

\bibitem{Hollik:2005ns}
W.~Hollik, U.~Meier, and S.~Uccirati,
\newblock Phys. Lett. {\bf B632}, 680 (2006), arXiv:hep-ph/0509302.

\bibitem{Hollik:2006ma}
W.~Hollik, U.~Meier and S.~Uccirati,
Nucl. Phys. B \textbf{765}, 154 (2007),
arXiv:hep-ph/0610312 

\bibitem{Awramik:2006ar}
M.~Awramik, M.~Czakon and A.~Freitas,
Phys. Lett. B \textbf{642} (2006), 563,
arXiv:hep-ph/0605339

\bibitem{Awramik:2006uz}
M.~Awramik, M.~Czakon, and A.~Freitas,
\newblock JHEP {\bf 0611}, 048 (2006), arXiv:hep-ph/0608099.

\bibitem{Dubovyk:2019szj}
I.~Dubovyk, A.~Freitas, J.~Gluza, T.~Riemann and J.~Usovitsch,
JHEP \textbf{08}, 113 (2019),
arXiv:1906.08815

\bibitem{Barbieri:1983wy}
R.~Barbieri and L.~Maiani,
\newblock Nucl. Phys. {\bf B224}, 32 (1983).


\bibitem{Grifols:1983gu}
J.~A.~Grifols and J.~Sola,
Phys. Lett. B \textbf{137}, 257 (1984)



\bibitem{Lim:1983re}
C.~S. Lim, T.~Inami, and N.~Sakai,
\newblock Phys. Rev. {\bf D29}, 1488 (1984).

\bibitem{Eliasson:1984yu}
E.~Eliasson,
\newblock Phys. Lett. {\bf B147}, 65 (1984).

\bibitem{Hioki:1985wz}
Z.~Hioki,
\newblock Prog. Theor. Phys. {\bf 73}, 1283 (1985).

\bibitem{Grifols:1984xs}
J.~A. Grifols and J.~Sola,
\newblock Nucl. Phys. {\bf B253}, 47 (1985).

\bibitem{Barbieri:1989dc}
R.~Barbieri, M.~Frigeni, F.~Giuliani, and H.~E. Haber,
\newblock Nucl. Phys. {\bf B341}, 309 (1990).

\bibitem{Drees:1990dx}
M.~Drees and K.~Hagiwara,
\newblock Phys. Rev. {\bf D42}, 1709 (1990).

\bibitem{Drees:1991zk}
M.~Drees, K.~Hagiwara, and A.~Yamada,
\newblock Phys. Rev. {\bf D45}, 1725 (1992).

\bibitem{Chankowski:1993eu}
P.~H. Chankowski {\em et~al.},
\newblock Nucl. Phys. {\bf B417}, 101 (1994).

\bibitem{Chankowski:1992er}
P.~H. Chankowski, S.~Pokorski, and J.~Rosiek,
\newblock Nucl. Phys. {\bf B423}, 437 (1994), arXiv:hep-ph/9303309.

\bibitem{Garcia:1993sb}
D.~Garcia and J.~Sola,
\newblock Mod. Phys. Lett. {\bf A9}, 211 (1994).

\bibitem{Dabelstein:1995ui}
A.~Dabelstein, W.~Hollik, and W.~Mosle,
\newblock {Z boson observables in the MSSM},
\newblock in {\em {Perspectives for electroweak interactions in e+ e-
  collisions. Proceedings, Ringberg Workshop, Tegernsee, Germany, February 5-8,
  1995}}, pp. 0345--362, 1995, arXiv:hep-ph/9506251.

\bibitem{Pierce:1996zz}
D.~M. Pierce, J.~A. Bagger, K.~T. Matchev, and R.-j. Zhang,
\newblock Nucl. Phys. {\bf B491}, 3 (1997), arXiv:hep-ph/9606211.

\bibitem{Heinemeyer:2006px}
S.~Heinemeyer, W.~Hollik, D.~Stockinger, A.~M. Weber, and G.~Weiglein,
\newblock JHEP {\bf 08}, 052 (2006), arXiv:hep-ph/0604147.

\bibitem{Djouadi:1996pa}
A.~Djouadi {\em et~al.},
\newblock Phys. Rev. Lett. {\bf 78}, 3626 (1997), arXiv:hep-ph/9612363.

\bibitem{Djouadi:1998sq}
A.~Djouadi {\em et~al.},
\newblock Phys. Rev. {\bf D57}, 4179 (1998), arXiv:hep-ph/9710438.


\bibitem{Heinemeyer:2002jq}
S.~Heinemeyer and G.~Weiglein,
\newblock JHEP {\bf 10}, 072 (2002), arXiv:hep-ph/0209305.

\bibitem{Haestier:2005ja}
J.~Haestier, S.~Heinemeyer, D.~Stockinger, and G.~Weiglein,
\newblock JHEP {\bf 12}, 027 (2005), arXiv:hep-ph/0508139.

\bibitem{Heinemeyer:2004gx}
S.~Heinemeyer, W.~Hollik, and G.~Weiglein,
\newblock Phys. Rept. {\bf 425}, 265 (2006), arXiv:hep-ph/0412214.

\bibitem{Heinemeyer:2007bw}
S.~Heinemeyer, W.~Hollik, A.~M. Weber, and G.~Weiglein,
\newblock JHEP {\bf 04}, 039 (2008), arXiv:0710.2972.

\bibitem{Heinemeyer:2013dia}
S.~Heinemeyer, W.~Hollik, G.~Weiglein, and L.~Zeune,
\newblock JHEP {\bf 12}, 084 (2013), arXiv:1311.1663.

\bibitem{Stal:2015zca}
O.~St{\aa}l, G.~Weiglein, and L.~Zeune,
\newblock JHEP {\bf 09}, 158 (2015), arXiv:1506.07465.

\bibitem{Bertolini:1985ia}
S.~Bertolini,
\newblock Nucl. Phys. {\bf B272}, 77 (1986).

\bibitem{Hollik:1986gg}
W.~Hollik,
\newblock Z. Phys. {\bf C32}, 291 (1986).

\bibitem{Hollik:1987fg}
W.~Hollik,
\newblock Z. Phys. {\bf C37}, 569 (1988).


\bibitem{Denner:1991ie}
A.~Denner, R.~J.~Guth, W.~Hollik and J.~H.~Kuhn,
Z. Phys. C \textbf{51}, 695 (1991)

\bibitem{Froggatt:1991qw}
C.~D. Froggatt, R.~G. Moorhouse, and I.~G. Knowles,
\newblock Phys. Rev. {\bf D45}, 2471 (1992).


\bibitem{Chankowski:1999ta}
P.~H. Chankowski, M.~Krawczyk, and J.~Zochowski,
\newblock Eur. Phys. J. {\bf C11}, 661 (1999), arXiv:hep-ph/9905436.

\bibitem{Grimus:2007if}
W.~Grimus, L.~Lavoura, O.~M. Ogreid, and P.~Osland,
\newblock J. Phys. {\bf G35}, 075001 (2008), arXiv:0711.4022.


\bibitem{LopezVal:2012zb}
D.~Lopez-Val and J.~Sola,
\newblock Eur. Phys. J. {\bf C73}, 2393 (2013), arXiv:1211.0311.


\bibitem{Broggio:2014mna}
A.~Broggio, E.~J. Chun, M.~Passera, K.~M. Patel, and S.~K. Vempati,
\newblock JHEP {\bf 11}, 058 (2014), arXiv:1409.3199.


\bibitem{Lu:2022bgw}
C.~T.~Lu, L.~Wu, Y.~Wu and B.~Zhu,
arXiv:2204.03796


\bibitem{Bahl:2022xzi}
H.~Bahl, J.~Braathen and G.~Weiglein,
arXiv:2204.05269


\bibitem{Babu:2022pdn}
K.~S.~Babu, S.~Jana and V.~P.~K.,
arXiv:2204.05303


\bibitem{Heo:2022dey}
Y.~Heo, D.~W.~Jung and J.~S.~Lee,
arXiv:2204.05728


\bibitem{Ahn:2022xeq}
Y.~H.~Ahn, S.~K.~Kang and R.~Ramos,
arXiv:2204.06485


\bibitem{Han:2022juu}
X.~F.~Han, F.~Wang, L.~Wang, J.~M.~Yang and Y.~Zhang,
arXiv:2204.06505


\bibitem{Arcadi:2022dmt}
G.~Arcadi and A.~Djouadi,
[arXiv:2204.08406]


\bibitem{Ghorbani:2022vtv}
K.~Ghorbani and P.~Ghorbani,
arXiv:2204.09001


\bibitem{Abouabid:2022lpg}
H.~Abouabid, A.~Arhrib, R.~Benbrik, M.~Krab and M.~Ouchemhou,
arXiv:2204.12018


\bibitem{Lee:2022gyf}
S.~Lee, K.~Cheung, J.~Kim, C.~T.~Lu and J.~Song,
arXiv:2204.10338


\bibitem{Benbrik:2022dja}
R.~Benbrik, M.~Boukidi and B.~Manaut,
arXiv:2204.11755


\bibitem{Botella:2022rte}
F.~J.~Botella, F.~Cornet-Gomez, C.~Mir\'o and M.~Nebot,
arXiv:2205.01115


\bibitem{Frandsen:2022xsz}
M.~T.~Frandsen and M.~Rosenlyst,
arXiv:2207.01465



\bibitem{Hessenberger:2016atw}
S.~Hessenberger and W.~Hollik,
Eur. Phys. J. C \textbf{77}, 178 (2017),
arXiv:1607.04610



\bibitem{Gunion:2002zf}
J.~F. Gunion and H.~E. Haber,
\newblock Phys. Rev. {\bf D67}, 075019 (2003), arXiv:hep-ph/0207010.

\bibitem{gunion:1990}
{J. F. Gunion, H. E. Haber, G. L. Kane and S. Dawson},
\newblock {\em {The Higgs hunter's guide}} (Perseus Publishing, Cambridge,
  Mass., 1990).

\bibitem{Bernon:2015qea}
J.~Bernon, J.~F. Gunion, H.~E. Haber, Y.~Jiang, and S.~Kraml,
\newblock Phys. Rev. {\bf D92}, 075004 (2015), arXiv:1507.00933.


\bibitem{Glashow:1976nt}
S.~L. Glashow and S.~Weinberg,
\newblock Phys. Rev. {\bf D15}, 1958 (1977).

\bibitem{Paschos:1976ay}
E.~A. Paschos,
\newblock Phys. Rev. {\bf D15}, 1966 (1977).



\bibitem{Branco:2011iw}
G.~C. Branco {\em et~al.},
\newblock Phys. Rept. {\bf 516}, 1 (2012), arXiv:1106.0034.

\bibitem{Deshpande:1977rw}
N.~G. Deshpande and E.~Ma,
\newblock Phys. Rev. {\bf D18}, 2574 (1978).

\bibitem{Ma:2006km}
E.~Ma,
\newblock Phys. Rev. {\bf D73}, 077301 (2006), arXiv:hep-ph/0601225.

\bibitem{Barbieri:2006dq}
R.~Barbieri, L.~J. Hall, and V.~S. Rychkov,
\newblock Phys. Rev. {\bf D74}, 015007 (2006), arXiv:hep-ph/0603188.

\bibitem{LopezHonorez:2006gr}
L.~Lopez~Honorez, E.~Nezri, J.~F. Oliver, and M.~H.~G. Tytgat,
\newblock JCAP {\bf 0702}, 028 (2007), arXiv:hep-ph/0612275.

\bibitem{Belyaev:2016lok}
A.~Belyaev, G.~Cacciapaglia, I.~P. Ivanov, F.~Rojas, and M.~Thomas,
\newblock (2016), arXiv:1612.00511.



\bibitem{Klimenko:1984qx}
K.~G. Klimenko,
\newblock Theor. Math. Phys. {\bf 62}, 58 (1985),
\newblock [Teor. Mat. Fiz.62,87(1985)].

\bibitem{Maniatis:2006fs}
M.~Maniatis, A.~von Manteuffel, O.~Nachtmann, and F.~Nagel,
\newblock Eur. Phys. J. {\bf C48}, 805 (2006), arXiv:hep-ph/0605184.

\bibitem{Staub:2017ktc}
F.~Staub,
\newblock Phys. Lett. {\bf B776}, 407 (2018), arXiv:1705.03677.

\bibitem{Lee:1977eg}
B.~W. Lee, C.~Quigg, and H.~B. Thacker,
\newblock Phys. Rev. {\bf D16}, 1519 (1977).

\bibitem{Lee:1977yc}
B.~W. Lee, C.~Quigg, and H.~B. Thacker,
\newblock Phys. Rev. Lett. {\bf 38}, 883 (1977).

\bibitem{Casalbuoni:1986hy}
R.~Casalbuoni, D.~Dominici, R.~Gatto, and C.~Giunti,
\newblock Phys. Lett. {\bf B178}, 235 (1986).

\bibitem{Casalbuoni:1987cz}
R.~Casalbuoni, D.~Dominici, F.~Feruglio, and R.~Gatto,
\newblock Nucl. Phys. {\bf B299}, 117 (1988).

\bibitem{Maalampi:1991fb}
J.~Maalampi, J.~Sirkka, and I.~Vilja,
\newblock Phys. Lett. {\bf B265}, 371 (1991).

\bibitem{Kanemura:1993hm}
S.~Kanemura, T.~Kubota, and E.~Takasugi,
\newblock Phys. Lett. {\bf B313}, 155 (1993), arXiv:hep-ph/9303263.

\bibitem{Akeroyd:2000wc}
A.~G. Akeroyd, A.~Arhrib, and E.-M. Naimi,
\newblock Phys. Lett. {\bf B490}, 119 (2000), arXiv:hep-ph/0006035.

\bibitem{Horejsi:2005da}
J.~Horejsi and M.~Kladiva,
\newblock Eur. Phys. J. {\bf C46}, 81 (2006), arXiv:hep-ph/0510154.

\bibitem{Ginzburg:2005dt}
I.~F. Ginzburg and I.~P. Ivanov,
\newblock Phys. Rev. {\bf D72}, 115010 (2005), arXiv:hep-ph/0508020.

\bibitem{Grinstein:2015rtl}
B.~Grinstein, C.~W. Murphy, and P.~Uttayarat,
\newblock JHEP {\bf 06}, 070 (2016), arXiv:1512.04567.

\bibitem{Cacchio:2016qyh}
V.~Cacchio, D.~Chowdhury, O.~Eberhardt, and C.~W. Murphy,
\newblock JHEP {\bf 11}, 026 (2016), arXiv:1609.01290.

\bibitem{Ginzburg:2010wa}
I.~F. Ginzburg, K.~A. Kanishev, M.~Krawczyk, and D.~Sokolowska,
\newblock Phys. Rev. {\bf D82}, 123533 (2010), arXiv:1009.4593.

\bibitem{Behrends:1955mb}
R.~E. Behrends, R.~J. Finkelstein, and A.~Sirlin,
\newblock Phys. Rev. {\bf 101}, 866 (1956).

\bibitem{Berman:1958ti}
S.~M. Berman,
\newblock Phys. Rev. {\bf 112}, 267 (1958).

\bibitem{Kinoshita:1958ru}
T.~Kinoshita and A.~Sirlin,
\newblock Phys. Rev. {\bf 113}, 1652 (1959).

\bibitem{vanRitbergen:1998yd}
T.~van Ritbergen and R.~G. Stuart,
\newblock Phys. Rev. Lett. {\bf 82}, 488 (1999), arXiv:hep-ph/9808283.

\bibitem{vanRitbergen:1999fi}
T.~van Ritbergen and R.~G. Stuart,
\newblock Nucl. Phys. {\bf B564}, 343 (2000), arXiv:hep-ph/9904240.

\bibitem{Steinhauser:1999bx}
M.~Steinhauser and T.~Seidensticker,
\newblock Phys. Lett. {\bf B467}, 271 (1999), arXiv:hep-ph/9909436.

\bibitem{Denner:1991kt}
A.~Denner,
\newblock Fortsch.Phys. {\bf 41}, 307 (1993), arXiv:0709.1075.

\bibitem{Hollik:1995dv}
W.~Hollik,
\newblock {Electroweak theory},
\newblock in {\em {5th Hellenic School and Workshops on Elementary Particle
  Physics, Corfu, Greece, September 3-24, 1995}},   arXiv:hep-ph/9602380.

\bibitem{Veltman:1977kh}
M.~J.~G. Veltman,
\newblock Nucl. Phys. {\bf B123}, 89 (1977).

\bibitem{Chanowitz:1978mv}
M.~S. Chanowitz, M.~A. Furman, and I.~Hinchliffe,
\newblock Nucl. Phys. {\bf B153}, 402 (1979).

\bibitem{Chanowitz:1978uj}
M.~S. Chanowitz, M.~A. Furman, and I.~Hinchliffe,
\newblock Phys. Lett. {\bf B78}, 285 (1978).


\bibitem{Marciano:1979yg}
W.~J. Marciano,
\newblock Phys. Rev. {\bf D20}, 274 (1979).

\bibitem{Sirlin:1983ys}
A.~Sirlin,
\newblock Phys. Rev. {\bf D29}, 89 (1984).



\bibitem{Weiglein:1998jz}
G.~Weiglein,
\newblock Acta Phys. Polon. {\bf B29}, 2735 (1998), arXiv:hep-ph/9807222.

\bibitem{Stremplat:1998}
A.~Stremplat,
\newblock Diploma thesis, Univ. of Karlsruhe, 1998.

\bibitem{Awramik:2003rn}
M.~Awramik, M.~Czakon, A.~Freitas, and G.~Weiglein,
\newblock Phys.Rev. {\bf D69}, 053006 (2004), arXiv:hep-ph/0311148.


\bibitem{Steinhauser:1998rq}
M.~Steinhauser,
\newblock Phys.Lett. {\bf B429}, 158 (1998), arXiv:hep-ph/9803313.

\bibitem{Sturm:2013uka}
C.~Sturm,
Nucl. Phys. B \textbf{874}, 698  (2013), 
arXiv:1305.0581


\bibitem{Jegerlehner:2001wq}
F.~Jegerlehner,
\newblock J.Phys. {\bf G29}, 101 (2003), arXiv:hep-ph/0104304.


\bibitem{Enomoto:2015wbn}
T.~Enomoto and R.~Watanabe,
\newblock JHEP {\bf 05}, 002 (2016), arXiv:1511.05066.


\bibitem{Hahn:2000kx}
T.~Hahn,
Comput. Phys. Commun. \textbf{140}, 418 (2001),
arXiv:hep-ph/0012260


\bibitem{Hahn:1998yk}
T.~Hahn and M.~Perez-Victoria,
Comput. Phys. Commun. \textbf{118}, 153 (1999),
arXiv:hep-ph/9807565 


\bibitem{Weiglein:1993hd}
G.~Weiglein, R.~Scharf and M.~Bohm,
Nucl. Phys. B \textbf{416}, 606 (1994),
arXiv:hep-ph/9310358

\bibitem{Weiglein:1992rj}
G.~Weiglein, R.~Mertig, R.~Scharf and M.~Bohm,
in
{\it New computing techniques in physics research II}.
Proceedings of the 2nd International Workshop on Software Engineering, 
Artificial Intelligence and Expert Systems in High-Energy and Nuclear Physics,
La Londe les Maures, France,  January 1992 (ed.~D.~Perret-Gallix),
World Scientific, Singapore 1992


\bibitem{Hahn:2015gaa}
T.~Hahn and S.~Pa\ss{}ehr,
Comput. Phys. Commun. \textbf{214}, 91-97 (2017),
arXiv:1508.00562


\bibitem{Heinemeyer:1998yj}
S.~Heinemeyer, W.~Hollik and G.~Weiglein,
Comput. Phys. Commun. \textbf{124}, 76 (2000),
arXiv:hep-ph/9812320 


\bibitem{Hahn:2009zz}
T.~Hahn, S.~Heinemeyer, W.~Hollik, H.~Rzehak and G.~Weiglein,
Comput. Phys. Commun. \textbf{180}, 1426 (2009),


\bibitem{Bahl:2018qog}
H.~Bahl, T.~Hahn, S.~Heinemeyer, W.~Hollik, S.~Pa\ss{}ehr, H.~Rzehak and G.~Weiglein,
Comput. Phys. Commun. \textbf{249}, 107099 (2020),
arXiv:1811.09073


\bibitem{THDMEWPOS}
https://github.com/st3v3m4n/THDM\_EWPOS





\end{thebibliography}

\end{document}